\documentclass[a4paper,11pt]{article}
\pdfoutput=1 
\usepackage{jcappub}

\usepackage[english]{babel}
\usepackage{amsmath}
\usepackage{array}
\usepackage{graphicx, xcolor}
\usepackage{ulem}
\usepackage{float}

\newcommand{\bx}{{\hbox{\boldmath $x$}}}
\newcommand{\bk}{{\hbox{\boldmath $k$}}}
\newcommand{\bq}{{\hbox{\boldmath $q$}}}
\newcommand{\bp}{{\hbox{\boldmath $p$}}}
\newcommand{\sbx}{{\hbox{\boldmath\scriptsize $x$}}}
\newcommand{\sbk}{{\hbox{\boldmath\scriptsize $k$}}}
\newcommand{\sbp}{{\hbox{\boldmath\scriptsize $p$}}}

\usepackage{simpler-wick}
\usepackage{comment}
\usepackage[bottom]{footmisc}
\begin{document}

\title{How Dilatation Invariance Suppresses Loop Corrections to Curvature Perturbations on CMB Scales}

\author[a,b]{Danilo Artigas,}
\emailAdd{artigas@tap.scphys.kyoto-u.ac.jp}
\affiliation[a]{
Department of Physics, Kyoto University, Kyoto 606-8502, Japan
}
\affiliation[b]{Theory Center, Institute of Particle and Nuclear Studies (IPNS),
High Energy Accelerator Research Organization (KEK),
Oho 1-1, Tsukuba 305-0801, Japan}

\author[c,d]{Shi Pi,}
\emailAdd{shi.pi@itp.ac.cn}
\affiliation[c]{
Institute of Theoretical Physics,
Chinese Academy of Sciences, Beijing 100190, China
}
\affiliation[d]{
Kavli Institute for the Physics and Mathematics of the Universe (WPI), UTIAS, 
The University of Tokyo, Kashiwa, Chiba 277-8583, Japan
}

\author[a,e]{Takahiro Tanaka}
\emailAdd{t.tanaka@tap.scphys.kyoto-u.ac.jp}
\affiliation[e]{
Center for Gravitational Physics and Quantum Information, Yukawa
Institute for Theoretical Physics, Kyoto University, Kyoto 606-8502, Japan
}

\author[b,f]{and Yuko Urakawa}
\emailAdd{yukour@post.kek.jp}
\affiliation[f]{The Graduate University for Advanced Studies (SOKENDAI),
Tsukuba 305-0801, Japan}

\abstract{
Departures from standard slow-roll (SR) inflation have attracted increasing interest in recent years. In particular, scenarios that strongly enhance the power spectrum of the curvature perturbation are often proposed as a mechanism for producing primordial black holes, which could account for part or the totality of dark matter. An ongoing debate in these models is whether non-linear interactions of cosmological perturbations on small scales, namely loop corrections, can become sufficiently large to backreact on CMB scales. 
In this work, we adopt a non-linear framework, dropping spatial-gradient terms, to study loop corrections at super-Hubble scales. Our focus is to clarify the role of spatial-diffeomorphism invariance, especially dilatation invariance, which is the symmetry under overall rescaling of spatial coordinates, in demonstrating the suppression of loop corrections to the CMB power spectrum. This analysis is valid for any inflaton potential, both with smooth and sharp transitions, and at any loop order, as long as the CMB scales cross the horizon during an SR phase. We compare this result with the different explanations proposed in the literature using various gauge choices, and show our analysis is consistent both in the comoving and in the spatially-flat gauges. In particular, we show that in the spatially-flat gauge, the cubic interaction contains a term that diverges in the sharp transition limit. However, this contribution is exactly cancelled by the quartic interaction.}
\maketitle

\section{Introduction} \label{sec:Intro}

Inflation provides a compelling mechanism for explaining the observed homogeneity, isotropy, and near scale invariance of the primordial perturbations that seed the large-scale structure of the Universe. In the simplest slow-roll (SR) scenario, the comoving curvature perturbation $\zeta$ is conserved outside the Hubble horizon, and primordial non-Gaussianities are constrained to be small. In recent years, however, growing interest has been devoted to departures from the SR evolution that can enhance the power spectrum of $\zeta$. A major motivation is that an enhancement of the power spectrum, from the value on cosmic-microwave-background (CMB) scales, $P_{\zeta\zeta}\left(k_{\text{CMB}}\right)\sim 10^{-9}$, to the value on smaller super-Hubble scales, $P_{\zeta\zeta}\left(p_c\right)\sim 10^{-2}$, could significantly increase the abundance of primordial black holes (PBHs), potentially accounting for the totality of dark matter \cite{Carr:2016drx,Carr:2020xqk,Green:2020jor,Escriva:2022duf,Byrnes:2025tji}.

A commonly studied mechanism for achieving such an enhancement is the ultra-slow-roll (USR) model, during which the inflaton potential remains approximately flat for a few $e$-folds \cite{Starobinsky:1992ts, Leach:2000yw, Leach:2001zf, Kinney:2005vj, Inoue:2001zt, Dimopoulos:2017ged,Pattison:2018bct,Pi:2022zxs,Escriva:2025ftp,Cruces:2026qvl}. In this phase, the curvature perturbation grows on super-Hubble scales as $d\zeta/dt \propto a^3$ until the required amplitude is reached. Since the resulting power spectrum remains smaller than unity, perturbation theory is generally expected to remain valid, although next-to-leading-order corrections may become non-negligible \cite{Martin:2012pe, Artigas:2024ajh, Caravano:2024moy, Caravano:2025diq}.

This expectation was put into question in Refs.~\cite{Kristiano:2022maq, Kristiano:2023scm}, where it was argued that the cubic self-interaction of the inflaton generates large, and possibly divergent, loop corrections which could backreact on CMB scales and ultimately lead to a breakdown of perturbation theory. This claim triggered an intense debate within the community. While some works continue to argue for potentially divergent loop corrections and the breakdown of perturbation theory \cite{Choudhury:2023vuj, Choudhury:2023jlt, Choudhury:2023rks, Choudhury:2023hvf, Kristiano:2024ngc, Kristiano:2024vst}, most studies instead conclude that loop corrections may indeed be enhanced, but nevertheless remain subdominant with respect to the tree-level contribution, thereby preserving the validity of perturbative expansion.

The relation among the different mechanisms responsible for this suppression, however, has not yet been made fully transparent. One line of argument concerns the sharpness of the transition into and out of the USR phase. Some studies claim that smoothing the transition regulates the divergent contributions \cite{Firouzjahi:2023aum, Firouzjahi:2023ahg}, whereas other works, including numerical analyses, suggest that loop corrections are small and comparable both for sharp and smooth transitions \cite{Cheng:2023ikq, Davies:2023hhn, Franciolini:2023agm}. Other proposed mechanisms involve the role of boundary terms in the action \cite{Fumagalli:2023zzl, Kawaguchi:2024lsw}, which can become non-negligible within the in-in formalism \cite{Arroja:2011yj}, the cancellation between cubic and quartic interaction terms \cite{Cheng:2023ikq, Maity:2023qzw, Iacconi:2023ggt, Inomata:2024lud, Ballesteros:2024zdp, Kawaguchi:2024rsv, Fumagalli:2024jzz}, and the backreaction terms which are manifest in the spatially-flat gauge \cite{Inomata:2025bqw, Inomata:2025pqa, Inomata:2026csq}. More general arguments based on consistency relations suggest that loop corrections on CMB scales must remain suppressed by virtue of residual symmetry constraints, even in the presence of a non-SR phase \cite{Riotto:2023hoz, Riotto:2023gpm, Motohashi:2023syh, Tada:2023rgp, Tasinato:2023ukp}. Related symmetry-based arguments have also been developed in terms of Ward identities or soft effective field theories, showing the absence of scale-invariant loop corrections in the strict soft limit \cite{Ema:2026dop}. See also \cite{Braglia:2025cee, Braglia:2025qrb, Braglia:2026fle} for approaches based on effective field theory.

Taken together, these works have already provided strong evidence that the large enhancement claimed in Refs.~\cite{Kristiano:2022maq, Kristiano:2023scm} does not survive once all relevant contributions are included. Nevertheless, it will still be useful to clarify how apparently large contributions arise in the middle of calculations and how they cancel in the final curvature perturbation. This is particularly important because many of the existing arguments are formulated either in a different gauge or in a form where the cancellation is shown not by directly calculating all the terms that appear in the spatially-flat gauge, but by using the consistency relation derived from the symmetry related to the spatial dilatation~\cite{Maldacena:2002vr}.

In this work, we adopt a non-linear framework, which can also be described by the $\delta N$ formalism \cite{Starobinsky:1982ee, Salopek:1990jq, Sasaki:1995aw, Sasaki:1998ug, Wands:2000dp, Lyth:2003im, Rigopoulos:2003ak, Lyth:2004gb, Lyth:2005fi, Artigas:2021zdk} (see also its extensions in \cite{Naruko:2012fe, Abolhasani:2013zya, Talebian-Ashkezari:2016llx, Talebian-Ashkezari:2018cax, Tanaka:2021dww,Tanaka:2023gul, Artigas:2024ajh,Ahmadi:2026rzf, Ahmadi:2026yrv}). This approach relies on a gradient expansion of the fields on super-Hubble scales. Within this framework, loop corrections are obtained through a perturbative expansion in which higher-order field fluctuations are determined iteratively from the linear solution. This procedure is the so-called Yang--Feldman formalism \cite{Yang:1950vi}. The $\delta N$ formalism has previously been employed to study loop corrections in \cite{Firouzjahi:2023ahg, Franciolini:2023agm, Iacconi:2023ggt, Inomata:2024lud, Inomata:2025bqw, Inomata:2025pqa, Iacconi:2026uzo, Iacconi:2026vyk,Inomata:2026csq}. The emphasis of our approach, however, differs in several respects.

First, we use the dilatation symmetry to identify the part of the long-wavelength curvature perturbation corresponding to Weinberg's adiabatic mode (WAM). In the gradient expansion, this symmetry, under a certain locality assumption~\cite{Tanaka:2017nff,Tanaka:2026zew}, implies that the non-linear evolution can be organised so that the WAM dependence drops out of the loop corrections to the final curvature perturbation. We then revisit the same statement without relying exclusively on the gradient expansion. In loop diagrams involving sub-horizon hard modes, the cancellation of the WAM contribution requires information about how the hard-sector state responds to the adiabatic mode. We show that, once this condition is satisfied, the WAM contribution is suppressed even when sub-horizon modes are fully taken into account.

This statement should be distinguished from the absence of all possible super-Hubble evolution of $\zeta$. In a non-attractor phase, the conjugate momentum of $\zeta$ can act as an independent soft variable. Its contribution is not fixed by the dilatation consistency relation alone, because the latter controls only the WAM generated by the residual dilatation. The momentum contribution is suppressed in the far-infrared limit if the soft mode has already been projected onto the attractor direction before the onset of the non-attractor phase. If this condition is not satisfied, however, the contribution associated with the conjugate momentum cannot be discarded on the basis of dilatation invariance alone. This distinction is useful for separating the cancellation of the WAM from possible effects of genuinely non-adiabatic soft variables.

Second, we trace the same cancellation in the spatially-flat gauge, which is the gauge used in the original calculation of Refs.~\cite{Kristiano:2022maq, Kristiano:2023scm}. In this gauge, individual contributions can exhibit the same type of enhancement that motivated the claim of a large loop correction. We show explicitly that these enhanced pieces cancel only after the non-linear relation to $\zeta$ and the relevant higher-order terms are consistently included. This provides a direct diagnostic of where the apparent enhancement can enter and why it does not survive in the final curvature perturbation.

This article is organised as follows. In Sec.~\ref{Sec:PertComovingGauge}, we derive the non-linear action at leading order in the SR parameters and obtain a solution for the comoving curvature perturbation. We show that the residual dilatation symmetry requires the WAM dependence of the non-linear solution to take a restricted form, which leads to the suppression of loop corrections on CMB scales in the gradient expansion. We also discuss how a perturbative expansion of the action may obscure this symmetry and artificially generate large intermediate contributions. In Sec.~\ref{Sec:omitted}, we revisit the cancellation from the viewpoint of the WAM. We show that the same cancellation persists in diagrams involving sub-horizon modes once the locality condition for the hard modes is imposed, and we distinguish the adiabatic WAM contribution from possible contributions associated with the conjugate momentum of $\zeta$. In Sec.~\ref{Sec:PertFlatGauge}, we present our study in the spatially-flat gauge and show how it is related to the comoving gauge analysis. We compute $\zeta$ explicitly up to third order in perturbations and show that, in this gauge, the divergence associated with the sharp transition in the cubic interaction is exactly cancelled by a corresponding contribution arising from the quartic interaction term and the non-linear relation between the gauges. Our conclusions are summarised in Sec.~\ref{Sec:Conclusion}.

\section{Perturbation in the comoving gauge} \label{Sec:PertComovingGauge}
In this section, for illustrative purposes, we first show the suppression of loop effects with external legs in the long-wavelength limit by considering only the leading order in the gradient expansion. In this approximation, all spatial gradients are neglected, and the modes on the CMB scale, whose power spectrum is our main focus, are assumed to cross the horizon during the SR phase. We should note that in this approximation, all relevant modes are outside the horizon scale from the beginning. To simplify the discussion, e.g.~to avoid entering into the issue of infrared divergences associated with primordial perturbations, we restrict our attention to a finite volume corresponding to our observable universe and assume that longer-wavelength perturbations can be neglected. Later, in Sec.~\ref{Sec:omitted}, we show that our conclusion remains unchanged even when the omitted contributions coming from the neglected sub-horizon modes are taken into account.

\subsection{Long-wavelength limit of the action during inflation}  \label{SSec:LW}
The action for the flat Friedmann universe whose matter content is 
a single scalar field $\phi$ with the potential $V(\phi)$ is given by 
\begin{align}
  S=\int d\xi \left[\frac1N \left(-\frac{3}{\kappa} a a'{}^2+\frac12 a^3 \phi'{}^2 \right)-Na^3 V(\phi)\right]\,,
  \label{Eq:strtingaction}
\end{align}
where $\kappa=8\pi G$ and the prime denotes the derivative with respect to the time coordinate $\xi$. 
We decompose the scale factor $a$ and the inflaton field $\phi$ into background plus perturbations as 
\begin{align}
    a=e^{\rho+\mathcal{R}}\,,\qquad \phi=\bar\phi+\varphi\,, 
\end{align}
where $\rho$ and $\bar\phi$ represent the background values, and $N$ is the lapse function. 
Notice that we neglect vector and tensor perturbations in the action, since these contributions are negligible at large scales, see App.~\ref{App:NonLocalTerms}. In this section, we choose the comoving slicing condition $\varphi=0$ as the gauge condition. 
Then, $\mathcal{R}$ in this gauge is the non-linearly defined so-called gauge-invariant comoving curvature perturbation, $\zeta$. 
Substituting $\varphi=0$ and $\mathcal{R}=\zeta$, we obtain 
\begin{align}
  S=\int d\xi\, e^{3\rho+3\zeta} 
   \biggl[&\frac{\rho'{}^2}N \biggl(-\frac{3}{\kappa} (1+\zeta_{,\rho})^2 +\frac12\bar\phi_{,\rho}^2 \biggr)
       -NV(\bar \phi)\biggr]\,. 
\end{align}
Solving the constraint equation, obtained by the variation with respect to $N$, we get 
\begin{align}
 N=\sqrt{\frac{\frac{3}{\kappa}\rho'{}^2(1+\zeta_{,\rho})^2-\frac12\rho'{}^2\bar\phi_{,\rho}^2}{V(\bar\phi)}}\,.
\end{align}
Substituting this expression for $N$ back into the action, 
and using the background $e$-folding number $\rho$ as the time coordinate instead of $\xi$, 
we obtain
\begin{align}
S=-2\int d\rho\, e^{3\rho+3\zeta}
\sqrt{V(\bar\phi)\left(\frac{3}{\kappa}{(1+\zeta_{,\rho})}^2-\frac12\bar\phi_{,\rho}^2\right)}~.
\label{eq:reduced_action}
\end{align}
Using the background Friedmann equation,
\begin{align} 
 \label{FriedmannEqn}
  H^2:=\frac{\rho'{}^2}{N^2}=\frac{\kappa V(\bar\phi)}{3-\epsilon_1}\,,
\end{align}
where 
\begin{align}
 \epsilon_1=-\frac{d\log H}{d\rho}=\frac{\kappa}{2}\bar\phi_{,\rho}^2\,,
\end{align} 
we have 
\begin{align}
 S=-\frac{2}{\kappa}\int d\rho\, e^{3\rho+3\zeta}H
\sqrt{\left(3(1+\zeta_{,\rho})^2-\epsilon_1\right)
  \left(3-\epsilon_1\right)}\,.
\label{action:generallong}
\end{align}
This is the complete reduced action in the long-wavelength limit in the comoving gauge. 
Of course, this action fully respects the dilatation invariance.\footnote{Under the coordinate rescaling $x^i \rightarrow \tilde{x}^i=x^i e^{-s}$, the curvature perturbation transforms as $\zeta\rightarrow \tilde{\zeta} = \zeta+s$, upon neglecting gradient terms, while its derivatives remain invariant. 
The total action then becomes
\begin{align}
 S= \int d^3\tilde{x} \, e^{3s} \int d\rho \, e^{3\rho + 3\tilde{\zeta} - 3s } L\left(\tilde{x}^i\right) \,, 
\end{align}
where $L$ denotes the Lagrangian of Eq.~\eqref{action:generallong} without the exponential factor, and is here taken at leading order in Taylor expansion around $\tilde{x}^i$. The action, therefore, remains invariant in this sense.
}

\subsection{Quasi-de Sitter expansion}
We assume $\epsilon_1\ll1$, although the other SR parameters
\begin{align}
    \epsilon_{i+1} &:= \partial_\rho \log \epsilon_i \quad , \quad \forall \, i\geq1\,,
\end{align}
may be large. In this approximation, we can expand \eqref{action:generallong} up to $\mathcal{O}\left(\epsilon_1\right)$ and use the integration by parts
\begin{align}
  \int d\rho\, e^{3\rho+3\zeta}3(1+\zeta_{,\rho}) H
   =& \int d\rho\, e^{3\rho+3\zeta} H \epsilon_1
   + \mbox{(surface terms)}\,,
\end{align}
to rewrite the action as
\begin{align}
 S= \frac1\kappa\int d\rho\, e^{3\rho+3\zeta}H \epsilon_1 
   \frac{\zeta_{,\rho}^2}{(1+\zeta_{,\rho})}+{\cal O}\left(\epsilon_1^2\right)\,. 
\label{eq:action_expansion}
\end{align}
This small $\epsilon_1$ expansion does not ruin the dilatation invariance. Hence, the terms lower than quadratic order in perturbation all vanish in Eq.~\eqref{eq:action_expansion}. 
At leading order in $\epsilon_1$, the Euler-Lagrange equation derived from the action~\eqref{eq:action_expansion} reads 
\begin{align} \label{EulerLagrangeZeta}
2\frac{\zeta_{,\rho\rho}}{1+\zeta_{,\rho}}
  +q(\zeta_{,\rho}^2+2\zeta_{,\rho})+3\zeta_{,\rho}^2=0\,, 
\end{align}
with 
\begin{align}
    q(\rho) := 3 + \frac{d\log\left(H\epsilon_1\right)}{d\rho} = 3 - \epsilon_1 + \epsilon_2\,.
\end{align}
As a result of the dilatation invariance, $\zeta$ without differentiation does not 
appear. 
Therefore, this equation is just a first-order differential equation for $\zeta_{,\rho}$, which has a solution of the form 
\begin{align}
 \zeta=\zeta_*+f(\zeta_{,\rho *})\,,
 \label{eq:zetaf} 
\end{align}
 where the subscript $*$ means that the values are evaluated at the initial time $\rho_*$. Here, the detailed form of the function $f$ is not important. The crucial point is that $\zeta_*$ only appears as a simple additive term, in general. 

As an example, let us restrict ourselves to a special USR evolution with $q=-3$. In this case, Eq.~\eqref{EulerLagrangeZeta} reduces to
\begin{equation}
 \zeta_{,\rho\rho}-3\zeta_{,\rho}(1+\zeta_{,\rho})=0,
\label{Eq:zetaeq}
\end{equation}
which yields the general solution, 
\begin{align}
\zeta=\zeta_*-\frac13\log\left[1-\zeta_{,\rho*}\left(e^{3(\rho-\rho_*)}-1\right)\right]\,.
\label{eq:USRsol}
\end{align}
Upon denoting $\psi$ the linear solution for $\zeta$, Eq.~\eqref{Eq:zetaeq} at linear level reduces to the superhorizon Mukhanov-Sasaki equation
\begin{equation}\label{eom:psi}
    \psi_{,\rho\rho}-3\psi_{,\rho}=0,
\end{equation}
whose solution~\eqref{PsiRhoLinSol} is now $\psi_{,\rho}=\zeta_{,\rho*}\, e^{3(\rho-\rho_*)}$. The non-linear solution for $\zeta$ given in \eqref{eq:USRsol} can therefore be rewritten as 
\begin{equation}\label{zeta(psi)}
    \zeta=\zeta_* -\frac13\log\left(1-\psi_{,\rho}+\zeta_{,\rho*}\right)\,.
\end{equation}
This expression does not contain an exponentially growing factor. 
In practically interesting cases, the initial condition can be determined from linear perturbations at some super-Hubble scale, and then we may rewrite the expression \eqref{zeta(psi)} using $\zeta_* = \psi_* $ and $\zeta_{,\rho*} = \psi_{,\rho*}$.

In general, the linearisation of Eq.~\eqref{EulerLagrangeZeta} admits the solution
\begin{align}
    \psi_{,\rho} = \zeta_{,\rho*} \,e^{-\int_{\rho_*}^\rho d\tilde{\rho} \,q(\tilde{\rho})}\,. \label{PsiRhoLinSol}
\end{align}
As a consequence, even if the non-linear function $f$ in Eq.~\eqref{eq:zetaf} may contain an exponentially large factor $\exp\left({-\int d\rho \, q}\right)$ for negative values of $q$, this factor would be absorbed by writing the solution in terms of the linear solution $\psi_{,\rho}$, instead of the initial value $\zeta_{,\rho*}$. Thus the general solution can therefore be rewritten as
\begin{align}
    \zeta = \zeta_* + g\left(\psi_{,\rho}\right) \,,
    \label{eq:generalForm}
\end{align}
and the non-linear function $g\left(\psi_{,\rho}\right)$ will not contain large numerical coefficients.

\subsection{Power spectrum on CMB scales}
We show that the loop corrections to the power spectrum on very large scales are generally suppressed, even if the non-linearity becomes large on a smaller scale. We assume $\psi$ and $\psi_{,\rho}$ to have Gaussian distribution and define the power spectra
\begin{align}
 P_{\psi\psi}(k)\delta^3(\bk+\bk')=k^3\langle \psi\left(\bk
 \right) \psi\left(\bk'
 \right)\rangle\,,
\end{align}
and $P_{\psi\psi_{,\rho}}$ and $P_{\psi_{,\rho}\psi_{,\rho}}$ are defined in the same way. 
In this section, we assume that $\psi(\bk)\approx \psi_*(\bk)$ on a CMB scale $k$, i.e.~the constant adiabatic mode dominates, at least at the linear level. During the SR phase, the time dependence of the mode exponentially decays. Therefore, even after a short period of the USR phase, the mode remains dominated by the constant mode. 

To compute the correlation functions, one may Taylor expand Eq.~\eqref{eq:generalForm} as follows:
\begin{align}
    \zeta = \psi_* + \sum_{m\geq 0} \frac{g^{(m)}(0)}{m!} \, \psi_{,\rho}^m \,, \label{ZetaTaylor}
\end{align}
where the superscript in parentheses $(m)$ denotes the $m$-th differentiation with respect to the argument, which is $\psi_{,\rho}$ here. When focusing on the two-point correlation function up to one-loop level, it is sufficient to expand Eq.~\eqref{eq:generalForm} up to third order. This yields\footnote{The disconnected diagrams such as
\begin{align}
    \langle \zeta_{\sbk} \rangle^2 = \left( \frac{g^{(2)}}{2} \int d^3q \,  \frac{P_{\psi_{,\rho} \psi_{,\rho}}(q)}{q^3} \delta^3(\bk) \right)^2\,,
\end{align}
do not contribute to our correlators, which should only contain connected diagrams.
}
\begin{align}
    \langle \zeta_{\sbk}\zeta_{\sbk'}\rangle
    =& \langle \psi \psi \rangle  + 2 \langle \psi \, g(\psi_{,\rho})\rangle + \langle g(\psi_{,\rho}) \,g(\psi_{,\rho})\rangle \cr
    =& \frac{P_{\psi\psi}(k)}{k^3} \delta^3\left(\bk+\bk'\right) \left[ 1 + 2 g^{(1)} \frac{P_{\psi\psi_{,\rho}}(k)}{P_{\psi\psi}(k)}  + \left(g^{(1)}\right)^2 \frac{P_{\psi_{,\rho} \psi_{,\rho}}(k)}{P_{\psi\psi}(k)}  \right. \label{zetazeta} \\
    & \left. + g^{(3)} \frac{P_{\psi\psi_{,\rho}}(k)}{P_{\psi\psi}(k)} \int d^3q \frac{P_{\psi_{,\rho} \psi_{,\rho}}(q)}{q^3}  + \frac{\left(g^{(2)}\right)^2}{2} \int d^3q \, \frac{k^3}{q^3 |\bk-\bq|^3} \frac{P_{\psi_{,\rho} \psi_{,\rho}}(q) P_{\psi_{,\rho} \psi_{,\rho}}(|\bk-\bq|)}{P_{\psi \psi}(k)} \right]  \,. \nonumber 
\end{align}
During the USR phase, because of the rapid growth of the time-dependent part of $\psi$ on super-horizon scales, although it remains much smaller than the constant part, one may think that these loop corrections might be enhanced, as claimed in Refs.~\cite{Kristiano:2022maq, Kristiano:2023scm}. However, when we impose that the perturbative expansion is valid at shorter wavelengths, the enhancement in the power spectrum on a large distance scale is suppressed, as shown below.

Let us denote by $k$ the co-moving wavenumber of a mode on a sufficiently large distance scale. Therefore, we can safely assume the following:  
\begin{align}
\frac{P_{\psi \psi_{,\rho}}(k)}{P_{\psi \psi}(k)} \ll 1 \quad , \quad \frac{P_{\psi_{,\rho} \psi_{,\rho}}(k)}{P_{\psi \psi}(k)} \ll 1 \,.
\label{eq:suppressionfactors}
\end{align}
Hence, we find that the terms proportional to $g^{(1)}$ or $(g^{(1)})^2$ in Eq.~\eqref{zetazeta} are negligible on the CMB scale. If it is not suppressed on the CMB scale, the same term will break the validity of perturbation on smaller scales because the suppression factors in \eqref{eq:suppressionfactors} are much larger (less suppressed) for a larger $k$.

Regarding the last term proportional to $(g^{(2)})^2$, 
one can use the estimate
\begin{align}
&\int d^3p \frac{k^3}{p^3 |\bp-\bk|^3} P_{\psi_{,\rho} \psi_{,\rho}}(p)P_{\psi_{,\rho} \psi_{,\rho}}(|\bp-\bk|) \approx C \frac{k^3}{p_c^3}\,, 
\end{align}
where $C$ is some constant and $p_c$ is the scale of the peak of $P_{\psi_{,\rho} \psi_{,\rho}}(p)$. 
This contribution is necessarily suppressed by the factor $k^3/p_c^3$ on super-horizon scales, unless the perturbative expansion breaks down on a smaller scale. As a result, we can conclude that the power spectrum is dominated by the first term in the square brackets in Eq.~\eqref{zetazeta}. 

On the other hand, the above arguments do not rule out large non-Gaussianities for shorter-wavelength modes that exit the horizon not much before the USR phase starts, contrary to what was claimed in Refs.~\cite{Kristiano:2022maq, Kristiano:2023scm}. This is obvious from the fact that the suppression of higher-order terms in the power spectrum on a CMB scale $k$ is based on the smallness of $\psi_{,\rho}(\bk)$ and $k^3/p_c^3$.
Since higher orders in Taylor expansion, Eq.~\eqref{ZetaTaylor}, only bring terms $\mathcal{O}\left(\psi_{,\rho}^m\right)$, the argument can easily be extended to any $n$-point correlation function and at any loop level.

\subsection{Errors caused by truncated calculations}
In many previous calculations, the action truncated at cubic order is used~\cite{Kristiano:2022maq, Kristiano:2023scm,Choudhury:2023vuj, Choudhury:2023jlt, Choudhury:2023rks, Choudhury:2023hvf, Kristiano:2024ngc, Kristiano:2024vst}. In this case, the dilatation invariance is violated, at least, in a straightforward computation. 
As a result, spurious behaviours appear in the IR dynamics. 
For illustrative purpose, we consider the action for $\zeta$ in the long-wavelength limit~\eqref{eq:action_expansion}, and truncate it at cubic order setting $q=-3$ to obtain
\begin{align}
 S^{\rm truncate} = \frac{1}{\kappa} \int d\rho\, e^{3\rho}H \epsilon_1 
   \zeta_{,\rho}^2(1-\zeta_{,\rho}+3\zeta)\,. \label{TruncatedAction} 
\end{align}
The variation of this action leads to the equation of motion, 
\begin{align}
  \zeta_{,\rho\rho}=3\zeta_{,\rho}\frac{1-2\zeta_{,\rho}+3\zeta}{1-3\zeta_{,\rho}+3\zeta}\,. 
  \label{eq:full_EOM}
\end{align}
This equation contains $\zeta$ without differentiation, which is the signature of the violation of the dilatation invariance. 
At this point, one may argue that since the action was truncated at third order in perturbations, the right-hand side of Eq.~\eqref{eq:full_EOM} should be expanded up to second order, and, as a consequence, the terms containing $\zeta$ would disappear. 
This, however, does not solve the problem in the ordinary field-theory path-integral calculations, in which we do not introduce any further artificial truncation. As a result, to avoid violating the dilatation invariance artificially, the full equation of motion \eqref{eq:full_EOM} should be solved without any further truncation. 
For illustrative purposese, we give a solution of this equation, but it is difficult to solve exactly. 
Hence, we iteratively solve it by rewriting it into the form,
\begin{align}
  \zeta_{,\rho\rho}=&3\zeta_{,\rho}\left(1+\Delta\right)\,,\cr
  \Delta:=&
     \zeta_{,\rho}+3\zeta_{,\rho}^2-3\zeta\zeta_{,\rho}+\cdots\,. 
\label{Eq:generalform}
\end{align}
At first order in $\Delta$, this equation can be solved as 
\begin{align}
 \zeta_{,\rho}=\zeta_{,\rho*}\exp\left(3(\rho-\rho_*) +3\int_{\rho_*}^\rho \Delta(\rho') \, d\rho'\right)\,.
\end{align}
Let us denote the long and short-wavelength part of $\zeta_{,\rho}$ by $\zeta_{,\rho}^{(L)}$ and 
$\zeta_{,\rho}^{(S)}$, respectively. 
Similarly, the linear perturbation variable $\psi$ is also decomposed into $\psi^{(L)}$ and $\psi^{(S)}$. 
Then, expanding the exponential, the third term of $\Delta$ brings a contribution to $\zeta_{,\rho}^{(L)}$, 
\begin{align}
\psi_{,\rho}^{(S)}(\rho) \int^\rho \psi^{(L)}(\rho')\psi_{,\rho}^{(S)}(\rho') \, d\rho' 
  \approx  3 \left(\psi^{(S)}\right)^2 \psi^{(L)} \,,
  \label{eq:cubiccontribution}
\end{align}
where we assume $\psi^{(L)}$ is dominated by the constant adiabatic mode, while $\psi^{(S)}$ is dominated by the non-constant mode which satisfies $\psi^{(S)}\approx\exp(3(\rho-\rho_*)) \psi^{(S)}_*$, i.e.~ $\psi_{,\rho}^{(S)}\approx 3\psi^{(S)}$ holds.
As a result, this term~\eqref{eq:cubiccontribution}
significantly contributes to the power spectrum 
of long-wavelength modes as a one-loop correction, 
even when $\psi^{(L)}$ is dominated by the adiabatic mode. 
This kind of fake enhancement is contained in Refs. \cite{Kristiano:2022maq,Kristiano:2023scm}. 
We stress again that the full expression, which respects the dilatation invariance, takes the form of \eqref{eq:generalForm} in general, and is free from such a spurious contribution. 

\subsection{Contributions from sub-horizon dynamics} \label{Sec:omitted} 
In this subsection, we discuss the contributions to long-wavelength perturbations from sub-horizon perturbations, which have been omitted so far. As discussed around Eq.~(\ref{eq:generalForm}), one can organise the influence of the non-linear evolution after $\rho=\rho_*$ in such a way that all $\psi$'s in the long-wavelength expression for $\zeta$ are associated with a time derivative, resulting in the suppression of loop corrections. This means that the subsequent time evolution in each local patch of the universe is independent of Weinberg's adiabatic mode (WAM) \cite{Weinberg:2003sw}.  In what follows, we show that this remains valid under an additional assumption about the locality of the quantum state of short modes \cite{Tanaka:2017nff,Tanaka:2026zew}, beyond the leading-order gradient expansion, i.e.~even when the contributions of sub-horizon modes are fully taken into account. 

We consider an inhomogeneous dilatation, which is parametrised by $s(\bx)$. We consider the situation in which $s(\bx)$ varies slowly on the scale of the long mode $k$. This inhomogeneous transformation is not a symmetry of the system, although it reduces to global dilatation symmetry when $s(\bx)$ is exactly constant. As formulated in Refs.~\cite{Tanaka:2017nff, Tanaka:2026zew}, when the system additionally satisfies a locality condition, the inhomogeneous dilatation can be treated as an approximate symmetry, in the sense that the action of this transformation acts on short modes in a rather trivial manner, as explained below. Then, the additive shift of $\zeta$ generated by the inhomogeneous dilatation transformation becomes an approximate solution of the system, which is the so-called WAM, i.e.~the well-known time-independent solution of $\zeta$ in the long-wavelength limit. The same locality condition also ensures the validity of the soft theorems for the WAM, namely the consistency relations in cosmology. We should recall that the consistency relations hold even when short modes are on sub-horizon scales. In what follows, we express the WAM generated by the inhomogeneous dilatation as $\zeta_{\rm ad}$ and the corresponding interaction-picture field as $\psi_{\rm ad}$.

Let us consider an operator which remains invariant under the inhomogeneous dilatation, which we denote by ${\cal O}^{({\rm gi})}(x)$.  
Since the WAM is the solution of $\zeta$ which is generated by the inhomogeneous dilatation transformation, by definition ${\cal O}^{({\rm gi})}(x)$ should be independent of the long-wavelength adiabatic perturbation $\psi_{\rm ad}^{(L)}$, that is
\begin{align}
 \frac{\partial \left\langle {\cal O}^{({\rm gi})}(\rho,\, \bx )\right\rangle_{\psi_{\rm ad}^{(L)}} }{\partial \psi^{(L)}_{\rm ad}(\bx)} =0\,, 
 \label{Def:gi}
\end{align}
where $\left\langle {\cal O}\right\rangle_{\psi_{\rm ad}^{(L)}}$ represents the expectation value of the operator ${\cal O}$ under the restriction to the fixed values of $\psi_{\rm ad}^{(L)}$. Based on this property, Ref.~\cite{Tanaka:2011aj} showed the suppression of the bispectrum in the squeezed limit in single-field SR inflation, and this result was subsequently reformulated in slightly different language in Ref.~\cite{Pajer:2013ana}. The locality condition implies
\begin{align}
 s(\bx)\frac{\partial \left\langle {\cal O}(\rho,\, \bx )\right\rangle_{\psi_{\rm ad}^{(L)}} }{\partial \psi^{(L)}_{\rm ad}(\bx)} \approx -\left\langle\delta_s {\cal O}\right\rangle_{\psi_{\rm ad}^{(L)}}\,,
 \label{Def:locality}
\end{align}
where $\delta_s$ is the inhomogeneous dilatation transformation with the parameter $s(\bx)$, and ${\cal O}$ is an arbitrary operator. Namely, the locality condition is the condition that the effects of the variation of the WAM are described by $\delta_s$. 
We take the expectation values on both sides, because the locality condition is a constraint on the (initial) quantum state. 

When one can neglect the correlations between other long-wavelength-mode operators and $\psi_{\rm ad}^{(L)}$, we find 
\begin{align}
 \langle \psi_{\rm ad}^{(L)} {\cal O}^{\rm (gi)}\rangle
  \approx \langle \psi_{\rm ad}^{(L)}\rangle
     \langle {\cal O}^{\rm (gi)}\rangle =0\,,
\label{eq:consistentcy_relation}
\end{align}
where the last equality follows simply from $\langle \psi_{\rm ad}^{(L)}\rangle=0$. 
On the other hand, from ${\cal O}$, an arbitrary short-wavelength operator which is gauge-invariant in an ordinary sense, we can construct an operator that satisfies $\delta_s {\cal O}^{({\rm gi})}(x)=0$ 
as\footnote{One example of an invariant operator under the inhomogeneous dilatation is the genuine gauge-invariant operator which was introduced in~\cite{Urakawa:2010it, Urakawa:2010kr} to discuss the infrared divergence of loops in the correlators of $\zeta$ and the gravitational waves in Refs.~\cite{Urakawa:2010it, Urakawa:2010kr, Tanaka:2013caa, Tanaka:2013xe, Tanaka:2014ina}. The genuine gauge invariance requires the invariance under a broader category of gauge transformations than the inhomogeneous dilatation. For our purpose in this paper, it is enough to consider a simplistic version of ${\cal O}^{({\rm gi})}(x)$, while it is not entirely free from the infrared enhancements.}
\begin{align}
    {\cal O}^{({\rm gi})}(\rho,\bx) \approx  {\cal O}(\rho, e^{-\zeta^{(L)}} \bx)\,.
    \label{Automatic_gi}
\end{align}
Expanding $ {\cal O}^{({\rm gi})}(\rho,\bx)$, we obtain
\begin{align}
{\cal O}^{({\rm gi})}(\rho,\bx)= 
{\cal O}(\rho, \bx)  - \zeta^{(L)}  \bx\frac{\partial {\cal O}}{\partial \bx} + \cdots  \,.
\label{exp:gi}
\end{align}
To show that substituting \eqref{exp:gi} into \eqref{eq:consistentcy_relation} reproduces the ordinary consistency relation, 
let us consider   
\begin{align}
{\cal O}_{\zeta\zeta,p}(\rho,\bx):=\int d^3k\, e^{i\sbk\cdot \sbx}
\zeta_{\sbp+\sbk/2}\zeta_{-\sbp+\sbk/2}=
\int \frac{d^3\Delta x}{(2\pi)^3}e^{-i\sbp\cdot \Delta\sbx} \zeta(\bx+\Delta\bx/2) \zeta(\bx-\Delta\bx/2)\,,  \label{eq:cr_derivation0}   
\end{align}
as an operator ${\cal O}$. 
This operator is a product of operators at two different points. Its gauge-invariant counterpart constructed following Eq.~\eqref{Automatic_gi} is expanded as\footnote{The integrand of Eq.~(\ref{eq:cr_derivation0}) can be made dilatation invariant by adding
\begin{align}
&\bigl[\zeta(\bx+\Delta\bx/2) \zeta(\bx-\Delta\bx/2)\bigr]^{\rm(gi)}  -
\left[\zeta(\bx+\Delta\bx/2) \zeta(\bx-\Delta\bx/2)\right]\cr
&\approx 
-\zeta^{(L)}\Biggl[\left\{\left(\bx+\frac{\Delta\bx}2\right)\frac{\partial}{\partial \bx}\zeta\left(\bx+\frac{\Delta\bx}{2}\right)\right\} 
\zeta\left(\bx-\frac{\Delta\bx}{2}\right)
+\zeta\left(\bx+\frac{\Delta\bx}{2}\right)
\left\{\left(\bx-\frac{\Delta\bx}2\right)\frac{\partial}{\partial \bx}
\zeta\left(\bx-\frac{\Delta\bx}{2}\right)\right\} 
\Biggr]
\cr
&=-\zeta^{(L)}
\left(\bx\frac{\partial}{\partial \bx}+\Delta\bx\frac{\partial}{\partial \Delta\bx}\right)\left\{\zeta\left(\bx+\frac{\Delta\bx}{2}\right) \zeta\left(\bx-\frac{\Delta\bx}{2}\right)\right\}\,,
\end{align}
which corresponds to the additional two terms in Eq.~(\ref{eq:cr_derivation}).} 
\begin{align}
{\cal O}^{\rm (gi)}_{\zeta\zeta,p}(\rho,\bx)
\approx {\cal O}_{\zeta\zeta,p}(\rho,\bx)
  +\zeta^{(L)}\left(\frac{\partial}{\partial \bp}\cdot\bp -\bx\cdot \frac{\partial}{\partial \bx}\right) 
  {\cal O}_{\zeta\zeta,p}(\rho,\bx)\,. \label{eq:cr_derivation}
\end{align}
Then, substituting this relation into Eq.~\eqref{eq:consistentcy_relation}, we obtain 
\begin{align}
 \langle \psi_{\rm ad}^{(L)}(\rho,\bx) {\cal O}_{\zeta\zeta,p}(\rho,\bx')\rangle \approx 
 - \langle \psi_{\rm ad}^{(L)}(\rho,\bx) \psi_{\rm ad}^{(L)}(\rho,\bx')\rangle  \frac{\partial}{\partial \bp}\cdot\bp \,\langle {\cal O}_{\zeta\zeta,p}(\rho,\bx')\rangle \,,  
\label{eq:consistentcy_relation2}
\end{align}
from which the ordinary consistency relation is derived by performing the Fourier transform with respect to $\bx-\bx'$ and replacing $\langle {\cal O}_{\zeta\zeta,p}(\rho,\bx)\rangle$ with the power spectrum $\sim |\zeta_p|^2$, which is independent of $\bx$.

Now, we use the locality condition to prove the absence of correlations between the leading-order long-wavelength curvature perturbation, $\psi^{(L)}$, and the higher-order corrections, $\delta \zeta^{(L)}:=\zeta^{(L)}-\psi^{(L)}$. 
The inhomogeneous dilatation transformations of  $\psi$ and $\zeta$ are given by 
\begin{align}
 \delta_s\psi^{(L)}=-s \left(1+\bx\cdot\frac{\partial  \psi^{(L)}}{\partial \bx} \right) \,,\qquad
 \delta_s\zeta^{(L)}=-s \left(1+\bx\cdot\frac{\partial \zeta^{(L)}}{\partial \bx} \right)\,.
\end{align}
Therefore, we have $\displaystyle\delta_s (\delta\zeta^{(L)})=-s\bx\cdot\frac{\partial (\delta\zeta^{(L)})}{\partial \bx}$. 
Thus, the locality condition \eqref{Def:locality} implies 
\begin{align}
    s \frac{\partial\left\langle \delta\zeta^{(L)}\right\rangle_{ \psi^{(L)}_{\rm ad}}}{\partial \psi^{(L)}_{\rm ad}} 
    \approx s \bx\cdot\frac{\partial}{\partial \bx}\left\langle\delta\zeta^{(L)} \right\rangle=0\,, 
\label{Eq:zeta3}
\end{align}
where the last equality holds because of the general translation invariance of the expectation values. 
Assuming that the correlations between $\psi^{(L)}_{\rm ad}$ and other long-wavelength perturbations are negligible, we find that this equality shows the suppression of the term $\langle \psi \zeta^{(3)} \rangle$, even after taking into account the loops including sub-horizon modes. The term $\langle \zeta^{(2)} \zeta^{(2)} \rangle$ is suppressed by $k^3$, for the same reason as in the case when sub-horizon modes are neglected.

While the main claim here is along the line of Refs.~\cite{Kawaguchi:2024rsv, Fumagalli:2024jzz}, there is a minor technical difference. While the contributions of the short modes are claimed to vanish in Refs.~\cite{Kawaguchi:2024rsv, Fumagalli:2024jzz}, since they can be summarised in a total derivative form such as
\begin{align}
-\int d\log k
 \frac{\partial P_{\zeta_{,\rho}\zeta_{,\rho}}(k)}{\partial \log k}\,.
 \label{eq:momentumIntegral}
\end{align}
if we introduce lower and upper cutoff momenta, $k_{\rm min}$ and $k_{\rm max}$, to the above momentum integral in this form, the boundary terms $\left[P_{\zeta_{,\rho}\zeta_{,\rho}}(k)\right]_{k_{\rm min}}^{k_{\rm max}}$ remain.
However, this form of momentum integral arises as a result of rewriting $\bx\cdot\frac{\partial}{\partial \bx}\left\langle\delta\zeta^{(L)} \right\rangle$ in Eq.~\eqref{Eq:zeta3} in terms of momentum variables. 
Let us take $(\zeta_{,\rho}(\rho,\bx))^2$ as an example of a term contained in $\delta \zeta^{(L)}$. In this case, $\langle \psi^{(L)} \zeta_{,\rho}(\rho,\bx))^2\rangle$ contains a factor
\begin{align}
\bx\cdot\frac{\partial}{\partial\bx}\langle (\zeta_{,\rho}(\rho,\bx))^2\rangle =&\, \bx\cdot\frac{\partial}{\partial\bx} 
  \int d^3 \bk\, d^3 \bk' \, e^{i(\sbk+\sbk')\cdot \sbx} 
  \langle \zeta_{,\rho\sbk}\zeta_{,\rho\sbk'}\rangle \cr 
  =&
  \int d^3 \bk \, d^3 \bk'\, \langle \zeta_{,\rho\sbk}\zeta_{,\rho\sbk'}\rangle   
\left(e^{i\sbk'\cdot \sbx} \bk\cdot\frac{\partial}{\partial\bk}e^{i\sbk\cdot \sbx} 
+e^{i\sbk\cdot \sbx} \bk'\cdot\frac{\partial}{\partial\bk'} e^{i\sbk'\cdot \sbx} \right)
\cr
  =&
  \int d^3 \bk \, \frac{P_{\zeta_{,\rho}\zeta_{,\rho}}(k)}{k^3}   
\bk\cdot\frac{\partial}{\partial\bk} 1\,,
\label{eq:momentumInt2}
\end{align}
which vanishes manifestly from the beginning because the expectation value $\langle (\zeta_{,\rho}(\rho,\bx))^2\rangle$ should be independent of $\bx$, and the expression in the last line
is also 0.  
We find that the expression in the last line formally coincides with \eqref{eq:momentumIntegral} by integrating by parts. 
Here, to justify this integration by parts, we may have to introduce step functions $\theta(k-k_{\rm min})\theta(k_{\rm max}-k)$ to represent the cutoff associated with $\langle \zeta_{,\rho\sbk}\zeta_{,\rho\sbk'}\rangle$. 
Hence, from the point of view that respects dilatation symmetry, there is no doubt about the claim that these step functions should be inside the differentiation $\partial/\partial\log k$ in the expression \eqref{eq:momentumIntegral}.

The above discussion does not, by itself, rule out contributions from terms involving $\psi^{(L)}$ that are not accompanied by derivative operators.
However, such contributions are typically suppressed by a factor $k^2$ or a larger power of $k$.
Hence, if it becomes significantly large on the CMB scales,
again the perturbative expansion breaks down on a smaller scale. Also, one may think that we should discuss correlation functions for the genuine gauge-invariant operator $\zeta^{{\rm (gi)}(L)}$. Here, for simplicity, we assume an IR cutoff beyond the largest scale of our observable universe, i.e.~the CMB scale. With this understanding, we can simply discuss the correlation functions of $\zeta^{(L)}$, as good proxies for observable correlation functions.

We should also note that the locality condition in the form of \eqref{Def:locality} or equivalently the consistency relation \eqref{eq:consistentcy_relation} is not guaranteed a priori. 
These relations require appropriate 
initial entanglement of short and long modes, which is a result of choosing the initial quantum state to respect the dilatation invariance. For example, the Euclidean vacuum, which reduces to the Bunch-Davies vacuum at the linear level, satisfies these conditions. Meanwhile, if we choose the quantum state, e.g.~to be the free-field vacuum at a finite initial time, there is obviously no short and long entanglement at the initial time. 
Then, the consistency relations such as~\eqref{eq:consistentcy_relation2} cannot be satisfied there in general. 

\section{Perturbation in the flat slicing} \label{Sec:PertFlatGauge}
Many previous calculations adopted the flat slicing as a gauge condition. 
In this gauge, the symmetry under the dilatation transformation is not manifest. 
Nevertheless, this gauge choice is convenient in SR single-field inflation models. However, it is not convenient in the USR inflation, as we shall clearly see below.

\subsection{Action in the long-wavelength limit}
The flat gauge is defined by setting $\mathcal{R}=0$. In this time-slicing, the gauge-invariant curvature perturbation $\zeta_n$ is defined by 
\begin{align}
  \zeta_n :=& -\frac{\varphi}  
  {\bar\phi_{,\rho}} \,, 
  \label{def:zetan}
\end{align}
which leads to 
\begin{align}
    \zeta_{n,\rho} =& - \frac{\varphi_{,\rho}}{\bar{\phi}_{,\rho}} + 3 \zeta_n  +  \frac{V_{,\phi}}{H^2 \bar{\phi}_{,\rho}} \zeta_n \,.
    \label{zetanrho}
\end{align}
We start by rewriting the effective action by setting $\mathcal{R}=0$ in Eq.~\eqref{Eq:strtingaction}, instead of setting $\varphi=0$. This yields
\begin{align}
 S=-2\int d\rho\, e^{3\rho}
\sqrt{V(\bar\phi+\varphi)\left(\frac{3}{\kappa}-
  \frac12{(\bar\phi_{,\rho}+\varphi_{,\rho})}^2\right)}\,.
\end{align}
Using the background Friedmann equation \eqref{FriedmannEqn}, the action can be rewritten as
\begin{align}
    S &= -2 \int d\rho\, e^{3\rho} \frac{\overline{V}}{H}\left[1+\frac{H^2}{2\overline{V}}\left(-\frac{1}{2}\varphi_{,\rho}^2 - \bar{\phi}_{,\rho}\varphi_{,\rho}\right)\right] \left[1+\frac{1}{2 \overline{V}}\left(\overline{V}^{(1)}\varphi + \frac{1}{2!}\overline{V}^{(2)}\varphi^2 + \cdots \right)\right]\,,
\end{align}
where we Taylor-expanded the potential around the background $\bar{\phi}$ and the square root, anticipating the  truncation at ${\cal O}(\epsilon_1)$. Here, the potential can be arbitrary. We denote by $\overline{V}^{(m)}$ the $m$-th derivative of ${V}(\phi)$ with respect to $\phi$ evaluated at $\phi=\bar\phi$. We also note that 
$\bar{\phi}_{,\rho} \propto \epsilon_1^{1/2}$ from the definition of $\epsilon_1$. Using Eqs.~\eqref{FriedmannEqn} and \eqref{def:zetan}, this implies that
\begin{align}
  \varphi \propto \epsilon_1^{1/2}\,,\qquad
  \overline{V}^{(m)}\propto \epsilon_1^{1-m/2}\,, \, \qquad \forall\, m\geq 1\,. 
\end{align}
Therefore, all terms $\overline{V}^{(m)} \varphi^m$ are of linear order in $\epsilon_1$ for $m\geq 1$. 
Furthermore, Eq.~\eqref{zetanrho} yields $\varphi_{,\rho}  \propto \epsilon_1^{1/2}$. 
Truncating the action at $\mathcal{O}\left(\epsilon_1\right)$, 
we obtain
\begin{align}
 \delta S= \int d\rho\, \frac{e^{3\rho}}{H}
   \biggl[ \frac{H^2}2\varphi_{,\rho}^2 - &\biggl(
      \frac{1}{2!} \overline{V}^{(2)} \varphi^2+\frac1{3!} \overline{V}^{(3)} \varphi^3     +\frac1{4!}\overline{V}^{(4)}\varphi^4+\cdots\biggr)\biggr]\,,
\end{align}
where we performed an integration by parts and used 
\begin{align}
    \int d\rho \, e^{3\rho} \left(\frac{\overline{V}^{(1)}\varphi}{2 H}- \frac{H}{2}\bar{\phi}_{,\rho} \varphi_{,\rho} \right) = 0 \,, 
\end{align}
which follows from the background equation, $\bar{\phi}_{,\rho\rho}=-3\bar{\phi}_{,\rho}-\overline{V}^{(1)}/H^2$. 
In this limit, the action is identical to the one of the perturbation of a test scalar field on a fixed de Sitter space, which is easy to handle. 
However, it contains second and higher derivatives of the potential. 
In a model with sharp transitions, such as the Starobinsky model, those derivatives are divergent 
(or still very large if the transition is smoothed out) at the transition. 
Therefore, the fact that terms in the action are $\propto \epsilon_1$ does not guarantee the smallness of each interaction term. 
For this reason, the use of the curvature perturbation in the flat gauge, $\zeta_n$, makes the computation very complicated. 

Using Eqs.~\eqref{def:zetan} and \eqref{zetanrho}, it is easy to rewrite the action in terms of gauge-invariant quantities,
\begin{align}
 \delta S=\int d\rho\, \frac{e^{3\rho}}{H}
   \Biggl[& \frac{H^2}2 \Bigg[\bar\phi_{,\rho}^2 \zeta_{n,\rho}^2 -2\left(3 \bar{\phi}_{,\rho}^2 + \frac{\overline{V}^{(1)} \bar{\phi}_{,\rho}}{H^2} \right) \zeta_n \zeta_{n,\rho} + \left(3  \bar{\phi}_{,\rho}^2 + \frac{\overline{V}^{(1)} \bar{\phi}_{,\rho}}{H^2} \right)^2 \zeta_n^2 \Bigg] \nonumber \\
   &- \sum_{m\geq 2} \left(- \zeta_n \bar{\phi}_{,\rho}\right)^m \frac{\overline{V}^{(m)}}{m!} \Biggr]\,,
 \label{zetanaction}
\end{align}
which can be further simplified by integrating by parts the term proportional to $\zeta_n \zeta_{n,\rho}$, to obtain
\begin{align}
 \delta S=&\int d\rho\, \frac{e^{3\rho}}{H}
   \Biggl[ \frac{H^2}2  \bar\phi_{,\rho}^2 \zeta_{n,\rho}^2  - \sum_{m\geq 3} \left(- \zeta_n \bar{\phi}_{,\rho}\right)^m \frac{\overline{V}^{(m)}}{m!} \Biggr]\,.
 \label{zetanaction}
\end{align}

Upon using the relation
\begin{align}
  \overline{V}^{(3)} \bar\phi_{,\rho}& =-\frac{H^2}{2}\epsilon_2\epsilon_3
\left(3+\epsilon_2+\epsilon_3+\epsilon_4\right)\,,  
\end{align}
at leading order in $\epsilon_1$, one can check that the cubic-order action matches Eq.~(3.9) in Maldacena's paper \cite{Maldacena:2002vr}.\footnote{To see this, note that
\begin{align}
   e^{3\rho} \overline{V}^{(3)} \bar{\phi}_{,\rho}^3
   &= - \frac{H^2}{\kappa} \partial_\rho \left(e^{3\rho} \epsilon_1 \partial_\rho \epsilon_2 \right) = - H^2 \partial_\rho \left(e^{3\rho} \bar{\phi}_{,\rho}^2 \partial_\rho  \left(\bar{\phi}_{,\rho\rho}/\bar{\phi}_{,\rho}\right) \right) \,.
\end{align}
Using then an integration by parts, we can write the cubic-order term in Eq.~\eqref{zetanaction} as
\begin{align}
    \delta S_3 &= \frac{1}{2} \int d\rho\, H e^{3\rho} \zeta_n^2 \zeta_{n,\rho} \bar{\phi}_{,\rho}^2 \partial_\rho \left(\frac{\bar{\phi}_{,\rho\rho}}{\bar{\phi}_\rho}\right) \,.
\end{align}
}

\subsection{One-loop correction}

\subsubsection{Arbitrary potential}

Let us consider the truncation of the action in terms of $\zeta_n$ at cubic order in perturbations. If we change variables back to $\zeta$, the missing term for the one-loop calculation is just the fourth-order term $\frac{1}{24}\overline{V}^{(4)} \bar\phi_{,\rho}^4\zeta_n^4$. 
This leads to a deviation from \eqref{eq:generalForm} in $\zeta$ -- represented by the term $\zeta_n^{\rm (3,4pt)}$ later -- which is badly divergent when considering a possible discontinuity in $\overline{V}^{(1)}$. 
In case the computation of the one-loop correction for the truncated action in $\zeta_n$ does not show this divergence, some terms proportional to the equation of motion should be thrown away in the action, which corresponds to taking a different choice of perturbation variable~\cite{Arroja:2011yj}. 

Here, we briefly show the computation of the one-loop correction in the flat gauge, using the Yang-Feldman formalism.
The operator equation for $\zeta_n$ obtained from the action \eqref{zetanaction} reads
\begin{align}
\partial_\rho \left(e^{3\rho} H \bar{\phi}_{,\rho}^2 \zeta_{n,\rho}\right)+
\frac{e^{3\rho} }{H} \sum_{m\geq 3} \frac{ \left(-\bar{\phi}_{,\rho}\right)^m \overline{V}^{(m)}}{(m-1)!}\zeta_n^{m-1} = 0 \,.  \label{ELeqn}
\end{align}
The solution is expanded as
\begin{align} \label{ZetanIterativeSolution}
 \zeta_n & =\psi +\zeta^{(2)}_n+\zeta_n^{(3,{\rm 3pt})}+\zeta_n^{(3,{\rm 4pt})}+\cdots\,,
\end{align} 
where $\psi$ obeys the linearisation of Eq.~\eqref{ELeqn},
\begin{align}
    \psi_{,\rho\rho} = \left(2\alpha + 3\right) \psi_{,\rho} \,, \label{EOMPsi}
\end{align}
with $\alpha:=\overline{V}^{(1)}/\left(H^2 \bar\phi_{,\rho}\right)$, and the second and third order terms are iteratively computed as
\begin{align}
\zeta^{(2)}_n =&\frac{1}{2H^2}\int d\rho\, \frac{e^{-3\rho}}{\bar\phi_{,\rho}^2}\int d\rho \, e^{3\rho} \bar\phi_{,\rho}^3
   \overline{V}^{(3)}\psi^2\,,\cr
\zeta^{(3,{\rm 3pt})}_n :=&\frac{1}{H^2}\int d\rho\, \frac{e^{-3\rho}}{\bar\phi_{,\rho}^2}\int d\rho\, e^{3\rho} \bar\phi_{,\rho}^3 
   \overline{V}^{(3)}\psi\zeta^{(2)}_n\,,\cr
\zeta^{(3,{\rm 4pt})}_n :=&-\frac1{6H^2}\int d\rho\, \frac{e^{-3\rho}}{\bar\phi_{,\rho}^2}\int d\rho\, e^{3\rho} \bar\phi_{,\rho}^4
   \overline{V}^{(4)}\psi^3\,. \label{eq:IntegralsZeta}
\end{align}
Here, we use a slightly abusive notation and denote by $\rho$ all the integration variables. The contributions of cubic order are divided into two parts $\zeta^{(3,{\rm 3pt})}_n$ and $\zeta^{(3,{\rm 4pt})}_n$, which are composed of three-point vertices and a four-point vertex, respectively. 
We show graphically the contribution of all terms to the different one-loop diagrams in App.~\ref{App:LoopDiagrams}.

In the case of a rapid transition from the USR phase to the SR phase, 
$\overline{V}^{(2)}$ or equivalently $\epsilon_3$ contains a sharp peak, which gives a term proportional to a Dirac delta function in the limit of the instantaneous transition. 
The higher-order derivatives of the potential all contain the derivatives of this delta-like function. 
In particular, evaluation of the integral of products of the derivatives of the potential should be treated carefully. Additionally, $\overline{V}^{(1)}$, $\bar{\phi}_{,\rho\rho}$ and $\zeta_{,\rho\rho}$ rapidly change at the transition. 

We perform integration by parts so as to remove all $\overline{V}^{(m)}$ with $m\geq 2$. 
Integrating by parts $\overline{V}^{(3)}$ twice, $\zeta^{(2)}_n$ becomes
\begin{align}
 \zeta^{(2)}_n =\frac{\alpha}{2} \psi^2 + \zeta^{(2,A)}_n+  \zeta^{(2,B)}_n\,, \label{zetan2ndorder}
\end{align}
with
\begin{align}
   \zeta_n^{(2,A)} :=& - 2 \int d\rho \, \alpha \psi \psi_{,\rho} \,, \\
   \zeta^{(2,B)}_n :=&  \int d\rho \frac{e^{-3\rho}}{\bar\phi_{,\rho}^2}\int d\rho\, e^{3\rho} \alpha (\bar\phi_{,\rho})^2(\psi_{,\rho})^2\,,
\end{align}
which is very concise and completely regular, even in the sharp transition limit. We denote with a subscript $J$ quantities evaluated at the USR-to-SR transition time $\rho_J$. 
The regularity of $\zeta^{(2)}_n$ is expected because the effect of the truncation of the action at cubic order appears only from third-order perturbations. 

Upon performing an integration by parts in $\zeta_n^{(3,\rm{4pt})}$ to eliminate $\overline{V}^{(4)}\bar\phi_{,\rho}$, we obtain
\begin{align}
    \zeta_n^{(3,\rm{4pt})} =& - \frac{1}{6H^2} \int d\rho \, \overline{V}^{(3)} \bar\phi_{,\rho} \psi^3 \nonumber \\
    &- \frac{1}{2H^4} \int d\rho \frac{e^{-3\rho}}{\bar \phi_{,\rho}^2} \int d\rho  \, \biggl[ e^{3\rho} \overline{V}^{(3)} \bar\phi_{,\rho}^2 \psi^2  \left(\overline{V}^{(1)}\psi + H^2 \bar\phi_{,\rho} \left(2\psi - \psi_{,\rho}\right)\right) \biggl] \,.
\label{eq:zeta3,4pt}
\end{align}
When substituting $\zeta_n^{(2)}$ in the expression of $\zeta_n^{(3,\rm{3pt})}$, one finds that the contribution coming from the first term in Eq.~\eqref{zetan2ndorder} is canceled by the term including $\overline{V}^{(3)}\overline{V}^{(1)}\psi^3$ in the expression of $\zeta_n^{(3,\rm{4pt})}$ given in the second line of Eq.~\eqref{eq:zeta3,4pt}. All remaining terms in $\zeta_n^{(3,\rm{3pt})}$ and $\zeta_n^{(3,\rm{4pt})}$ can then be integrated by parts until all $\overline{V}^{(m)}$ with $m\geq 2$ are eliminated. As a consequence, all terms remain finite. We show the explicit expressions of $\zeta_n^{(3,\rm{3pt})}$ and $\zeta_n^{(3,\rm{4pt})}$ obtained after integrations by parts in App.~\ref{App:IPP}.

\subsubsection{Application: ultra-slow-roll transition}
To provide a concrete example, we consider a model potential with a step-function transition. Then, all non-linear interactions vanish in each flat region of the potential. 

To solve the long-wavelength evolution along the last SR phase, we fix the initial condition just before the USR-SR transition,  $\rho_J^- $. However, we abuse notation and use $\rho_J$ to represent the value evaluated at $\rho_J^+$, just after the transition. The background-field derivative evolves as follows:
\begin{align}
    \frac{\bar{\phi}_{,\rho}}{\bar{\phi}_{,\rho J}} = - \frac{\alpha_J}{3}  + \left(1 + \frac{\alpha_J}{3} \right) e^{-3(\rho-\rho_J)}\,.
\end{align}
Using the definition of $\epsilon_2$, this allows us to write
\begin{align}
     \alpha = & -\frac{3(\epsilon_{2J}+6)}{\epsilon_{2J}+6-\epsilon_{2J}\,e^{-3(\rho-\rho_J)} }\quad , \quad \alpha_J = - \frac{1}{2}\left(\epsilon_{2 J} + 6 \right)\,.
\end{align}
The linearised equation \eqref{EOMPsi} can be rewritten as
\begin{align}
    \varphi_{,\rho\rho} + 3\varphi_{,\rho} = - \frac{\overline{V}^{(2)} \varphi}{H^2} \,.
    \label{eq:varphieq}
\end{align}
Fixing the initial condition for $\varphi$ at $\rho_J^+$, and setting the potential second derivative on the right-hand side to be zero,  Eq.~\eqref{eq:varphieq} is solved as 
\begin{align}
    \varphi(\rho) = \varphi_J + \frac{1}{3}\varphi_{,\rho J}\left(1- e^{-3(\rho-\rho_J)}\right)\,. 
\end{align}
Using definitions \eqref{def:zetan} and \eqref{zetanrho} and the background Klein-Gordon equation, one can write
\begin{align}
    \frac{\varphi_{,\rho J}}{\bar{\phi}_{,\rho J}} = \left(\alpha_J +3\right) \psi_{J} - \psi_{,\rho J} \,.
\end{align}
Combining this expression with the previous equations for $\varphi$ and $\bar{\phi}_{,\rho}$, we obtain
\begin{align}
    \psi(\rho) =&  \psi_{J} + \frac{1- e^{-3(\rho-\rho_J)}}{-\alpha_J + (3+\alpha_J)e^{-3(\rho-\rho_J)}} \psi_{,\rho J} \,, \cr
    =& \psi_{J}+\frac{2\left(1-e^{-3(\rho-\rho_J)}\right)}{\epsilon_{2J}+6-\epsilon_{2J} \, e^{-3(\rho-\rho_J)}} \psi_{,\rho J}\,.
    \label{eq:solpsi}
\end{align}
Since the equation for $\psi$ does not have a term containing the Dirac's delta function ($q$ does not contain $\epsilon_3$ and higher-order SR parameters), $\psi_J$ and $\psi_{,\rho J}$ in Eq.\eqref{eq:solpsi} do not have any jump at $\rho_J$, and hence they can be evaluated at $\rho_J^-$, before the transition. Also, the solution \eqref{eq:solpsi} is valid from $\rho_J^-$. 
One can then compute the higher-order corrections of $\zeta_n$. At second order, the integrals in Eq.~\eqref{zetan2ndorder} can be performed assuming that $\overline{V}^{(1)}=0$ during the USR phase. Since non-linearities in $\zeta_n$ can be neglected during the USR phase, the integrals that determine the non-linear corrections can be started from $\rho_J^- $, just before the transition. The third-order contribution can be obtained by evaluating the integrals from $\rho_J^- $, as well. In the current model, the derivatives of the potential are given by
\begin{align}
    \overline{V}^{(1)}=& - \frac{H^2}{2} \left(\epsilon_{2J} + 6\right) \bar{\phi}_{,\rho J} \, \theta\left(\rho-\rho_J\right) \,, \\
        \overline{V}^{(2)}\bar{\phi}_{,\rho} =& - \frac{H^2}{2} \left(\epsilon_{2J} + 6\right) \bar{\phi}_{,\rho J} \, \delta\left(\rho-\rho_J\right) \,. 
\end{align}
Using the property
\begin{align}
    \theta\left(\rho-\rho_J\right) \, \delta\left(\rho-\rho_J\right) =& \frac{1}{2} \delta\left(\rho-\rho_J\right) \,,
\end{align}
the second order yields 
\begin{align}
\zeta^{(2)}_n
=&  -\left.\frac32\psi ^2+\psi\psi_{,\rho}
\right\vert_{\rho_J}\,,   
\label{eq:secondorder}
\end{align}
when evaluated at late times $\rho-\rho_J\gg 1$. 

The third-order perturbations can be evaluated similarly, using $\theta^2\left(\rho-\rho_J\right) \, \delta\left(\rho-\rho_J\right) = \frac{1}{3} \delta\left(\rho-\rho_J\right)$. 
Here, we introduce the regularised expressions $\tilde{\zeta}_n^{(3,\rm{3pt})}$ and $\tilde{\zeta}_n^{(3,\rm{4pt})}$ obtained from ${\zeta}^{(3,\mathrm{3pt})}_n$ and ${\zeta}^{(3,\mathrm{4pt})}_n$ 
subtracting the diverging terms shown in Eqs.~\eqref{eq:AppC}, which implicitly contain $\delta^2(\rho-\rho_J)$. The subtracted pieces exactly cancel between ${\zeta}^{(3,\mathrm{3pt})}_n$ and ${\zeta}^{(3,\mathrm{4pt})}_n$. Integrating by parts Eq.~\eqref{eq:ZetaTilde3ptNotSimplified} only once gives us an integrand proportional to $\overline{V}^{(2)}$. Since $\zeta_n^{(2,A)}$, $\zeta_n^{(2,B)}$ and $\zeta_{n,\rho}^{(2,B)}$ vanish at $\rho_J$, the expression for $\tilde{\zeta}_n^{(3,\rm{3pt})}$ can be reduced to
\begin{align}
    \tilde{\zeta}_n^{(3,\rm{3pt})} = &  \frac{2}{H^4} \int_{\rho_J^-}^{\infty} d\rho \, \frac{e^{-3\rho}}{\bar{\phi}_{,\rho}^2} \int_{\rho_J^-}^\rho d\rho \, \overline{V}^{(2)} \overline{V}^{(1)} e^{3\rho} \bar{\phi}_{,\rho} \psi \,, \\
    = & \left. \frac{1}{2}\left(\epsilon_{2J} + 6\right)\, \psi^2 \psi_{,\rho} \right\vert_{\rho_J} \,.
\end{align}
The expression for $\tilde{\zeta}_n^{(3,\rm{4pt})}$ can be computed similarly from Eqs.~\eqref{eq:ZetaTilde4ptNotSimplified1}, \eqref{eq:ZetaTilde4ptNotSimplified2} -- \eqref{eq:ZetaTilde4ptNotSimplified4}, reducing to
\begin{align}
     \tilde{\zeta}_n^{(3,\rm{4pt})} = & \left. 3\psi^3 - \frac{1}{2} \left(\epsilon_{2J} + 12 \right) \psi^2 \psi_{,\rho} + \psi \psi_{,\rho}^2 \right\vert_{\rho_J} \,.
\end{align}
Combining the two contributions, we finally obtain
\begin{align}
    \zeta_n = \left.\psi -\frac32\psi^2+\psi\psi_{,\rho}
+3\psi^3
-3\psi^2\psi_{,\rho}
 +\psi\psi_{,\rho}^2\cdots\right\vert_{\rho_J}\,, 
 \label{zetanfin}
\end{align}
as the asymptotic value at late times in the SR phase where $\zeta \approx \zeta_n$.

We should stress that the cancellation of the terms proportional to $\epsilon_{2J}$ in Eq.~\eqref{zetanfin} does not mean that there is no enhancement of perturbation. 
As $\psi$ and $\psi_{,\rho}$ continue to grow during the USR phase 
in proportion to $e^{3\rho}$, $\psi_{J}$ and $\psi_{,\rho J}$ are exponentially large. 
The important point is that the result is completely reproduced by the non-linear transformation from $\zeta_n$ to $\zeta$, as we shall see in the next subsection. During the SR phase succeeding the USR phase, $\zeta$ is conserved, while $\zeta_n$ varies to settle to the same value at a late epoch.

\subsection{Conversion between $\zeta_n$ and $\zeta$}

Let us assume here the potential to be flat $V_{,\phi} = 0$, for simplicity. We perform a non-linear gauge transformation from the flat gauge $\mathcal{R}=0$ to the comoving gauge where the scalar-field perturbations vanish,
\begin{align}
        \begin{cases}
        \mathcal{R}\left(\rho\right)=0\,, \\
               \phi\left(\rho\right)=\bar{\phi}\left(\rho\right)+\varphi \left(\rho\right)\,,               
            \end{cases} \rightarrow \quad
            \begin{cases}
               \zeta\left(\tilde{\rho}\right) = \tilde{\mathcal{R}}\left(\tilde{\rho}\right)= \rho - \tilde{\rho}\,, \\              \tilde{\phi}\left(\tilde{\rho}\right)=\bar{\phi}\left(\tilde{\rho}\right) =\phi\left(\rho\right)\,,
            \end{cases} 
\end{align}
where we associate variables in the comoving gauge with a tilde, e.g.~$\tilde{\mathcal{R}}$, $\tilde{\phi}$. According to the $\delta N$ formalism, perturbed long-wavelength variables also follow the background evolution equation.  
The scalar-field dynamics is governed by the Klein-Gordon equation $\phi_{,\rho\rho} = - 3 \phi_{,\rho}$. After solving this equation setting the initial value at $\tilde{\rho}$, we have
\begin{align}
 \bar{\phi}(\tilde{\rho}) =\phi(\rho)=& (\bar\phi(\tilde{\rho})+\varphi(\tilde{\rho}))
      -\frac13\left(\bar{\phi}_{,\rho}(\tilde{\rho})+\varphi_{,\rho}(\tilde{\rho})\right)\left(e^{-3{\zeta}}-1\right)\,,
\end{align}
which can be inverted as follows:
\begin{align}
 \zeta=\frac13\left\{
  \log\left[1+3\psi-\psi_{,\rho}\right]-\log\left[1-\psi_{,\rho}\right]
     \right\}\,,
     \label{eq:zetafull}
\end{align}
where we used the definitions of $\zeta_n$ and its derivative $\zeta_{n,\rho}$ given in Eqs.~\eqref{def:zetan} and\eqref{zetanrho}, and $\zeta_n$ is replaced with $\psi$, since they are identical for a flat potential, $V_{,\phi}=0$. From the equation of motion~\eqref{eom:psi}, one can see that the combination $3\psi-\psi_{,\rho}$ does not grow even during the USR phase. Hence, setting an initial condition during the flat-potential region, we can expand the first logarithm linearly and find that 
 \begin{align}
 \zeta\approx \psi_* - \frac13 \psi_{,\rho*} - \frac13 \log\left[1-\psi_{,\rho}\right]
     \,, 
     \label{eq:zeta38}
\end{align}
where the quantities with $*$ are the initial values. 
Equation \eqref{eq:zeta38} takes the form of $\zeta \approx \psi_{*}+g(\psi_{,\rho})$ as given in Eq.~\eqref{eq:generalForm}. Furthermore, as $\psi_{,\rho*}$ is small, it can be approximately rewritten as $\log[1+\psi_{,\rho*}]$, and we obtain $\zeta_{,\rho *}=\psi_{,\rho *}/(1-\psi_{,\rho *})$ from Eq.~\eqref{eq:zeta38}. Substituting these relations into Eq.~\eqref{zeta(psi)}, we find it agrees with Eq.~\eqref{eq:zeta38}.
If $\psi$ has already been sufficiently enhanced at $\rho_*$, $\psi_{,\rho}\approx 3\psi$ is satisfied including at $\rho_*$ and one recovers a well-known approximate formula $\zeta\approx -\ln\left[1-3\psi\right]/3$, as expected~\cite{Biagetti:2018pjj,Atal:2019cdz,Pi:2022ysn}.

Although $\psi$ grows exponentially during the USR phase, $\varphi$ remains small, and the linear perturbation theory remains valid. When evaluating \eqref{eq:zetafull} at $\rho_J$, at the end of the USR region, and expanding the expression up to third order in perturbation, one finds\footnote{Note that the same procedure can be performed for $V_{,\phi}=\text{const}$. In this case, the scalar field's equation of motion reads
\begin{align}
 \bar{\phi}(\tilde{\rho}) =& (\bar\phi(\tilde{\rho})+\varphi(\tilde{\rho}))
     - \frac{V_{,\phi}}{3H^2} \zeta -\frac13\left(\bar{\phi}_{,\rho}(\tilde{\rho})+\varphi_{,\rho}(\tilde{\rho}) + \frac{V_{,\phi}}{3 H^2}\right)\left(e^{-3{\zeta}}-1\right)\,,
\end{align}
which can be inverted to obtain 
\begin{align}
    \zeta =& \frac{3 H^2}{V_{,\phi}} \left[ \varphi - \frac{1}{3} \left(\bar{\phi}_{,\rho} + \varphi_{,\rho} + \frac{V_{,\phi}}{3 H^2}\right) \sum_{m\geq 1} \frac{(-3 \zeta)^m}{m!}  \right]\,,
    \label{NLzeta}
\end{align}
where we also expanded the exponential function. This equation can then be solved iteratively at each order in perturbation. At the third order, one finds
 \begin{align}
    \zeta =& \zeta_n + \frac{1}{4} \epsilon_2 \zeta_n^2 +  \zeta_n \zeta_{n,\rho} -\frac{\epsilon_2}{2} \zeta_n^3 +  \left[-\frac{3}{2} + \frac{\epsilon_2}{4}\right] \zeta_n^2 \zeta_{n,\rho}  + \zeta_n \zeta_{n,\rho}^2  \,,
\end{align}
where we recall that the second SR parameter is generally defined at leading order in $\mathcal{O}\left(\epsilon_1\right)$ as $\epsilon_2 = -6 - 2 V_{,\phi}/(H^2 \bar\phi_{,\rho})$. In the USR case, $\epsilon_2=-6$ and one recovers Eq.~\eqref{eq:FieldRedef}.
}
\begin{align}
    \zeta =& \left. \psi - \frac{3}{2} \psi^2 +  \psi \psi_{,\rho} + 3 \psi^3 - 3 \psi^2 \psi_{,\rho}  + \psi \psi_{,\rho}^2+\cdots \right|_{\rho_J} \,.\label{eq:FieldRedef}
\end{align}
This is consistent with the result obtained in Eq.~\eqref{zetanfin} and shows that $\zeta\approx\zeta_n$ at a late time in the subsequent SR phase. This was expected since in the SR phase the difference between $\zeta$ and $\zeta_n$ appears only at higher order in the SR expansion \cite{Malik:2008im}. 

\subsection{Boundary term}

Several works emphasise the importance of the boundary term for the computation of the loop correction \cite{Fumagalli:2023zzl, Kawaguchi:2024lsw}. Plugging the field redefinition Eq.~\eqref{eq:FieldRedef} into the truncated action in the comoving gauge Eq.~\eqref{TruncatedAction}, we can write contributions at second and third order in perturbations as
\begin{align}
    \delta S = \int d\rho \, L\left(\zeta\right) \approx \int d\rho \left[ L\left(\zeta_n\right) + \frac{\delta L}{\delta \zeta} \delta \zeta\left(\zeta_n, \zeta_{n,\rho}\right) \right]+(\mbox{boundary terms})\,.
    \label{eq:zetan_action}
\end{align}
The term proportional to the equation of motion contains higher-derivative terms. Removing these terms with an integration by parts, we arrive at the action in the flat gauge~\eqref{zetanaction}. The boundary terms in \eqref{eq:zetan_action} are necessary so as to cancel the boundary term after integration by parts. We should note that, if we neglect the boundary term, what one calculates as $\zeta_n$ is actually not $\zeta_n$ but $\zeta$, since the cubic interaction term proportional to $\delta L/\delta \zeta$ in Eq.~\eqref{eq:zetan_action} vanishes when we substitute the interaction-picture field. If we do not neglect the boundary terms, they take care of the necessary field redefinition~\cite{Arroja:2011yj}.

\section{Conclusion} \label{Sec:Conclusion}
In this work, we have investigated the behaviour of loop corrections to the comoving curvature perturbation during hypothetical deviations from the SR inflationary evolution. To analyse a possible breakdown of perturbation theory as claimed by different studies, we made use of a simple approximation at leading order of gradient expansion to demonstrate the importance of respecting the dilatation invariance, i.e.~the invariance under the rescaling of the spatial coordinates, in describing non-linear evolution in the super-Hubble regime.

We demonstrated how dilatation invariance ensures the suppression of loop corrections on sufficiently large scales, and how its violation can lead to an enhancement of loop corrections.
To be more precise, by solving the dynamics of the comoving curvature perturbation $\zeta$ on super-Hubble scales, we showed that dilatation invariance implies that non-linear corrections depend only on the initial velocity $\zeta_{,\rho*}$, see Eq.~\eqref{eq:zetaf}. As a consequence, loop corrections involve the power spectra $P_{\psi \psi_{,\rho}}$ and $P_{\psi_{,\rho} \psi_{,\rho}}$, where $\psi$ is the interaction-picture field. 
Such contributions can be enhanced during a non-SR phase, such as a USR phase, where the non-adiabatic mode grows. 
However, on CMB scales, these corrections are generally suppressed when compared to the tree-level power spectrum $P_{\psi \psi}$, since the non-adiabatic modes continue to decay before the non-SR phase starts, see \cite{Maity:2023qzw} for concrete estimations. Our result applies to an arbitrary potential shape, provided that the CMB-scale modes crossed the horizon during the first slow-roll phase, and can be straightforwardly generalised to higher-loop orders.

In several works, the action written in terms of $\zeta$ is truncated at third order in perturbations. We showed that this truncation breaks the dilatation invariance at the one-loop level and, therefore, yields spurious loop corrections of order $P_{\psi \psi}$, which are no longer suppressed on CMB scales, as observed in \cite{Kristiano:2022maq, Kristiano:2023scm,Choudhury:2023vuj, Choudhury:2023jlt, Choudhury:2023rks, Choudhury:2023hvf, Kristiano:2024ngc, Kristiano:2024vst}. 
We argue that such contributions should not appear in any consistent treatment that respects dilatation invariance. This is the underlying reason why the introduction of quartic interactions eliminates spurious loop corrections, as has already been explored in previous studies \cite{Cheng:2023ikq, Maity:2023qzw, Iacconi:2023ggt, Inomata:2024lud, Ballesteros:2024zdp, Kawaguchi:2024rsv, Fumagalli:2024jzz}.

To reconcile the different loop calculations in the comoving gauge and in the spatially-flat gauge, we have also studied the latter gauge. When using the field redefinition to rewrite the action in the flat gauge, the expression differs from the action in the comoving gauge by a boundary term only~\cite{Fumagalli:2023zzl, Kawaguchi:2024lsw}. If we carelessly neglect this boundary term, one may therefore be implicitly computing the curvature perturbation in flat gauge $\zeta$ instead of $\zeta_n$ and vice versa. Since $\zeta$ and $\zeta_n$ are non-linearly related to each other, this type of mistake can lead to a wrong estimate of the loop corrections.

After deriving the non-linear equation of motion for $\zeta_n$ at large scales, we used the Yang-Feldman formalism to solve the dynamical evolution iteratively up to third order in perturbations. In this way, we showed that the cubic and quartic interactions depend on the third and fourth derivatives of the potential, respectively. In models with a sharp transition such as the often-considered SR-USR-SR model, these interaction vertices are singular like the first and second derivatives of Dirac's delta function, and the one-loop corrections to the power spectrum originating from these vertices are both divergent, which echoes the results of Kristiano and Yokoyama. 
However, we showed that this divergence originating from the cubic interaction is exactly cancelled by an analogous term coming from the quartic interaction, as also shown by other studies \cite{Cheng:2023ikq, Maity:2023qzw, Iacconi:2023ggt, Inomata:2024lud, Ballesteros:2024zdp, Kawaguchi:2024rsv, Fumagalli:2024jzz}. We finally applied this analysis to the specific SR-USR-SR model and showed that the expression for $\zeta_n$ evaluated at a late time matches the one obtained from the non-linear evolution of $\zeta$, further showing the consistency of our results in the two different gauges.

We also briefly discussed the general suppression of loop corrections originating from short-wavelength modes, including sub-Hubble perturbations, although our main focus in this paper is on a concise description of the super-Hubble dynamics. We did not perform any detailed explicit calculations, but the absence of large loop corrections to the long-wavelength curvature perturbation from sub-Hubble modes follows from the requirement of dilatation invariance together with the locality condition for the quantum state of hard modes, which is equivalent to the so-called consistency relations in cosmological perturbation. 
This aspect is examined in more detail in Ref.~\cite{Tanaka:2026zew}. 

Finally, we mention that our explicit analysis relies on the basic idea of the standard separate-universe approach or the $\delta N$ formalism \cite{Starobinsky:1982ee, Salopek:1990jq, Sasaki:1995aw, Sasaki:1998ug, Wands:2000dp, Lyth:2003im, Rigopoulos:2003ak, Lyth:2004gb, Lyth:2005fi, Artigas:2021zdk,Cruces:2025typ}. This approach should be a good approximation in our case, since we are mainly interested in perturbations on the CMB scale. 
It is well known that, if the non-adiabatic mode dominates the adiabatic one, as it is the case for perturbations on relatively short-wavelength scales, around the peak of the power spectrum, the $\delta N$ formalism can break down, requiring higher orders in the gradient expansion~\cite{Naruko:2012fe, Artigas:2024ajh, Ahmadi:2026rzf, Ahmadi:2026yrv, Pattison:2019hef, Jackson:2023obv, Briaud:2025ayt}.

\section*{Acknowledgement}
We thank J.~Fumagalli, R.~Namba and L.~Pinol for discussions and useful comments. 
This work is supported in part by the National Key Research and Development Program of China Grant No.~ 2021YFC2203004.
D.~A.~thanks the long-term-visiting-fellow program of the High Energy Accelerator Research Organization (KEK), during which part of this project was conducted. D.~A.~was supported by JSPS Grant-in-Aid for Scientific Research Grant No.~JP23KF0247 during part of this work.
S.~P.~is supported by National Natural Science Foundation of China Grants No.~12475066 and No.~12447101, by JSPS KAKENHI Grant No.~JP24K00624, and by the World Premier International Research Center Initiative (WPI Initiative), MEXT, Japan.
T.~T.~is supported by Grant-in-Aid for Scientific Research under Contract No.~JP23H00110, and also by SPIRIT2 2026 of Kyoto University.
Y.~U.~is supported by Grant-in-Aid for Scientific Research under Contract No.~JP26H00402(26H00402),  Grant-in-Aid for Scientific Research (B) under Contract No.~JP23K25873 (JP23H01177), and JST FOREST Program under Contract No.~JPMJFR222Y.
\appendix
\section{Effective action and non-local terms} \label{App:NonLocalTerms}
In general, the reduced action written in terms of the master variables contains non-local terms. Here, we neglected such terms by starting with the homogeneous and isotropic metric ansatz, Eq.~\eqref{Eq:strtingaction}. Below, we show that this approximation is valid, because non-local terms are $\mathcal{O}\left(\epsilon_1^2\right)$. 

At the starting point~\eqref{Eq:strtingaction}, we did not consider the anisotropy of the metric. 
The 3-metric should be $a^2 \exp(h_{ij})$ with $h^j_j=0$. 
Here, the index is raised using $\delta^{ij}$. 
The traceless part of the extrinsic curvature arising from $h_{ij}$ is given by 
\begin{align}
 \tilde K_i^j=\frac1{2N}\exp(-h^{kj})\partial_\xi \exp(h_{ik})\,.
\end{align}
Using this quantity, the additional contribution missing in Eq.~\eqref{Eq:strtingaction} turns out to be  
\begin{align}
  \int \frac{d\xi}{2\kappa N} a^3\tilde K_i^{\,j} \tilde K_j^{\,i}\,.
\end{align}
If we reduce the action in the comoving gauge ($\varphi=0$), assuming this additional contribution is 
small, we find that Eq.~\eqref{eq:action_expansion} in the end gains an additional contribution 
\begin{align}
 \frac1{2\kappa}\int d\rho\, e^{3\rho+3\zeta}H 
 \frac{e^{-h^{ik}}e^{-h^{jl}}\partial_\rho e^{h_{ij}}\partial_\rho e^{h_{kl}}}{1+\zeta_{,\rho}}\,.
 \label{eq:addition_term}
\end{align}
If we do not neglect the spatial derivatives, the traceless part of the extrinsic curvature is written as 
\begin{align}
\tilde K_{ij}=\frac1{2N}\left[\partial_\xi \exp(h_{ik})-D_i N_j-D_j N_i\right]^{\rm TL}\,, 
\end{align}
where $N_i$ is the shift vector and ``TL'' represents the operation to take the traceless part. 
Here, we neglect the tensor perturbations. Then, the momentum constraint obtained from the variation with respect to $N_i$ with the aid of appropriate spatial gauge conditions tells us that $\tilde K_{ij}$ should be $O(\epsilon_1)$, and therefore the additional contribution given in Eq.~\eqref{eq:addition_term} is $O(\epsilon_1^2)$, which we can safely neglect within the analysis presented in this paper. 
The canonical conjugate of $h_{ij}$ before the phase-space reduction using the constraints and gauge-fixing conditions, 
\begin{align}
 e^{h_{ik}} \tilde\pi^{kj} =\frac{e^{3\rho+3\zeta}}{\kappa}H\, e^{-h^{jk}} {\tilde K_{ki}}\,,
\end{align}
becomes constant in the long-wavelength limit, and is essentially $Q_i^{\,j}$ introduced in Ref.~\cite{Tanaka:2021dww}. 
Constancy of $Q_i^{\,j}$ means that ${\tilde K_{kl}}$ decays approximately $\propto e^{-3\rho}$. 
Hence, even without the argument of the $O(\epsilon_1^2)$ suppression mentioned above, this term gives only a negligible contribution to the long-wavelength dynamics well outside the horizon. 

\section{Loop diagrams} \label{App:LoopDiagrams}
To compute the two-point function $\langle\zeta_n\zeta_n\rangle$ up to one loop, it is sufficient to expand $\zeta_n$ up to third order in perturbations, Eq.~\eqref{ZetanIterativeSolution}. We show below the diagrams of the different contributions to one loop. First, we describe $\psi:=\zeta_n^{(1)}$, $\zeta_n^{(2)}$, $\zeta_n^{\rm (3,3pt)}$ and $\zeta_n^{\rm (3,4pt)}$ by the diagrams
\begin{figure}[H]
\vspace*{-.5cm}
\begin{minipage}[c]{0.45\linewidth}
     \centering 
      \includegraphics[page=1, scale=1]{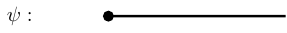}
\end{minipage},
    \hspace*{0cm}
\begin{minipage}[c]{0.45\linewidth}
     \centering 
     \includegraphics[page=2, scale=1]{Figures.pdf}
\end{minipage},\\
\begin{minipage}[c]{0.45\linewidth}
     \centering 
     \includegraphics[page=3, scale=1]{Figures.pdf},
\end{minipage},
    \hspace*{0cm}
\begin{minipage}[c]{0.45\linewidth}
    \centering
    \includegraphics[page=4, scale=1]{Figures.pdf}
\end{minipage}.
\end{figure}
The right ends of each diagram are associated with an interaction-picture field, $\psi$. 
When we compute the two-point function, two diagrams among them are multiplied, and pairs of $\psi$ are replaced with $\langle \psi\psi \rangle$. While the cancellation can be confirmed before taking the contraction, let us also show the resulting possible diagrams as follows:
\begin{figure}[H]
 \begin{minipage}[c]{0.45\linewidth}
     \centering 
    \includegraphics[page=5, scale=1]{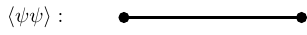}
\end{minipage},
\vspace*{5pt}
\begin{minipage}[c]{0.45\linewidth}
     \centering 
     \includegraphics[page=6, scale=1]{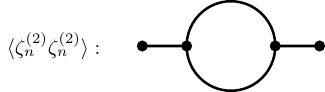} 
\end{minipage},\\[.5cm]
\hspace*{0pt}
 \begin{minipage}[c]{0.45\linewidth}
      \centering 
  \includegraphics[page=7, scale=1]{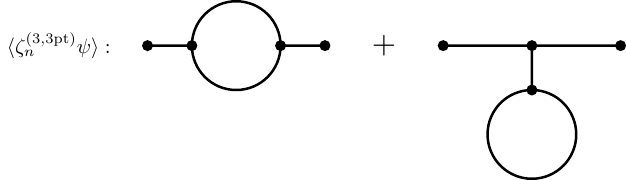} 
\end{minipage}
\hspace*{5cm} \raisebox{.8cm}{,}\\[.5cm]
\hspace*{-20pt}
 \begin{minipage}[c]{0.45\linewidth}
    \centering 
     \vspace*{-30pt}\includegraphics[page=8, scale=1]{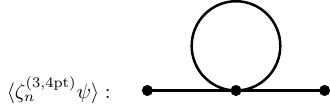}
\end{minipage}.
\end{figure}
\noindent Note that the rightmost diagram of $\langle\zeta_n^{(3,3\text{pt})}\psi\rangle$ can generally be removed by a background/tadpole renormalisation \cite{Senatore:2009cf,Kristiano:2025ajj}.

\section{Integration by parts of the third-order perturbations of $\zeta_n$} \label{App:IPP}
We denote by $\tilde \zeta^{(3,{\rm 3pt})}_n$ and $\tilde \zeta^{(3,{\rm 4pt})}_n$, respectively, the terms $ \zeta^{(3,{\rm 3pt})}_n$ and $ \zeta^{(3,{\rm 4pt})}_n$ from which the diverging term that mutually cancels has been subtracted. More precisely, $\tilde \zeta^{(3,{\rm 3pt})}_n + \tilde \zeta^{(3,{\rm 4pt})}_n = \zeta^{(3,{\rm 3pt})}_n + \zeta^{(3,{\rm 4pt})}_n $ and
\begin{align}
    \tilde \zeta^{(3,{\rm 3pt})}_n &\equiv  \zeta^{(3,{\rm 3pt})}_n - \frac{1}{2H^4} \int d\rho \frac{e^{-3\rho}}{\bar \phi_{,\rho}^2} \int d\rho  \,  e^{3\rho} \overline{V}^{(1)} \overline{V}^{(3)} \bar\phi_{,\rho}^2 \psi^3  \,,  \cr
        \tilde \zeta^{(3,{\rm 4pt})}_n &\equiv  \zeta^{(3,{\rm 4pt})}_n + \frac{1}{2H^4} \int d\rho \frac{e^{-3\rho}}{\bar \phi_{,\rho}^2} \int d\rho  \,  e^{3\rho} \overline{V}^{(1)} \overline{V}^{(3)} \bar\phi_{,\rho}^2 \psi^3  \,.
\label{eq:AppC}
\end{align}
In this section, we will show that $\tilde \zeta^{(3,{\rm 3pt})}_n$ and $\tilde \zeta^{(3,{\rm 4pt})}_n$ are indeed regular.

\subsection{Analysis of $\tilde \zeta^{(3,{\rm 3pt})}_n$}
\noindent By definition,
\begin{align}
    \tilde \zeta^{(3,{\rm 3pt})}_n &=   \frac{1}{H^2} \int d\rho \frac{e^{-3\rho}}{\bar \phi_{,\rho}^2} \int d\rho  \,  e^{3\rho} \overline{V}^{(3)} \bar\phi_{,\rho}^3 \psi \left(\zeta_n^{(2,A)} + \zeta_n^{(2,B)}\right) \,. \label{eq:ZetaTilde3ptNotSimplified}
\end{align}
Integrating this by parts to eliminate $\bar\phi_{,\rho} \overline{V}^{(3)}$ one finds
\begin{align}
    \tilde \zeta^{(3,{\rm 3pt})}_n &=  I^{(3,{\rm 3pt})} + J^{(3,{\rm 3pt})} + K^{(3,{\rm 3pt})} + L^{(3,{\rm 3pt})} \,,
\end{align}
where we defined
\begin{align}
    I^{(3,{\rm 3pt})} &\equiv \frac{1}{H^2} \int d\rho \, \overline{V}^{(2)} \psi \left(\zeta_n^{(2,A)} + \zeta_n^{(2,B)}\right) \,,  \\
    J^{(3,{\rm 3pt})} &\equiv  \frac{1}{H^2} \int d\rho  \frac{e^{-3\rho}}{\bar \phi_{,\rho}^2} \int d\rho  \,  e^{3\rho} \overline{V}^{(2)} \bar\phi_{,\rho}^2 \left(3\psi-\psi_{,\rho}\right) \left(\zeta_n^{(2,A)} + \zeta_n^{(2,B)}\right)\,,  \\
    K^{(3,{\rm 3pt})} &\equiv  \frac{2}{H^4} \int d\rho \frac{e^{-3\rho}}{\bar \phi_{,\rho}^2} \int d\rho  \,  e^{3\rho} \overline{V}^{(1)} \overline{V}^{(2)} \bar\phi_{,\rho} \psi \left(\psi \psi_{,\rho} + \zeta_n^{(2,A)} + \zeta_n^{(2,B)}\right) \,,  \\
    L^{(3,{\rm 3pt})} &\equiv - \frac{1}{H^4} \int d\rho \frac{e^{-3\rho}}{\bar \phi_{,\rho}^2} \int d\rho  \, \overline{V}^{(2)} \psi \int d\rho \, e^{3\rho} \overline{V}^{(1)} \bar\phi_{,\rho} \psi_{,\rho}^2 \,.    
\end{align}
For $I^{(3,{\rm 3pt})}$, $J^{(3,{\rm 3pt})}$ and $L^{(3,{\rm 3pt})}$ we further integrate by parts the term $\bar\phi_{,\rho} \overline{V}^{(2)}$ while for $K^{(3,{\rm 3pt})}$ we integrate by parts the term $2 \bar\phi_{,\rho} \overline{V}^{(1)} \overline{V}^{(2)}$. One then notices that all four terms are regular. By adding them altogether, we find the final expression
\begin{align}
    \tilde \zeta^{(3,{\rm 3pt})}_n = & \alpha \psi \left(\zeta_n^{(2,A)} + \zeta_n^{(2,B)}\right) + \int d\rho \, \alpha \psi_{,\rho} \left[ 3 \alpha \psi^2 - 2 \left(\zeta_n^{(2,A)} + \zeta_n^{(2,B)}\right)  \right] \nonumber \\
    & - \int d\rho \, \frac{e^{-3\rho}}{\bar \phi_{,\rho}^2} \left[ 2 \alpha \psi \int d\rho \, e^{3\rho} \alpha \bar\phi_{,\rho}^2 \psi_{,\rho}^2 + 3 \int d\rho \, e^{3\rho} \alpha^2 \bar\phi_{,\rho}^2 \psi \psi_{,\rho}^2 \right] \nonumber \\
    &+ 2 \int d\rho \, \frac{e^{-3\rho}}{\bar\phi_{,\rho}^2} \int d\rho \, \alpha \psi_{,\rho} \int d\rho \, e^{3\rho} \alpha \bar\phi_{,\rho}^2 \psi_{,\rho}^2 \,. \label{eq:ZetaTilde33pt}
\end{align}
It is then clear that the diagram is regular.

\subsection{Analysis of $\tilde \zeta^{(3,{\rm 4pt})}_n$}
\noindent We proceed similarly with
\begin{align}
    \tilde \zeta^{(3,{\rm 4pt})}_n =  - \frac{1}{6 H^2} \int d\rho \, \overline{V}^{(3)} \bar \phi_{,\rho} \psi^3 - \frac{1}{2 H^2} \int d\rho \, \frac{e^{-3\rho}}{\bar{\phi_{,\rho}^2}} \int d\rho \, e^{3\rho} \overline{V}^{(3)} \bar \phi_{,\rho}^3 \psi^2 \left(2\psi - \psi_{,\rho}\right) \,. \label{TildeZeta4}
\end{align}\\
\\
{\bf{First term:}}\\
The first term in the right-hand side of Eq.~\eqref{TildeZeta4} can be integrated by parts to obtain
\begin{align}
    - \frac{1}{6 H^2} \int d\rho \, \overline{V}^{(3)} \bar \phi_{,\rho} \psi^3 = - \frac{\overline{V}^{(2)} \psi^3}{6 H^2} + \frac{1}{2 H^2} \int d\rho \, \overline{V}^{(2)} \psi^2 \psi_{,\rho}\,,
\end{align}
After performing another integration by parts, we get
\begin{align}
  -  \frac{1}{6 H^2} \int d\rho \, \overline{V}^{(3)} \bar \phi_{,\rho} \psi^3 =  - \frac{\overline{V}^{(2)} \psi^3}{6 H^2} + \frac{1}{2} \alpha \psi^2 \psi_{,\rho} - \frac{1}{2} \int d\rho \, \alpha \psi \psi_{,\rho} \left[ \left(3 \alpha+ 6\right) \psi + 2 \psi_{,\rho} \right]   \,, \label{eq:ZetaTilde4ptNotSimplified1}
\end{align}
which remains regular when evaluated after the transition.\\
\\
{\bf{Second term:}}\\
The second term in the right-hand side of Eq.~\eqref{TildeZeta4} can also be integrated by parts to obtain
\begin{align}
    - \frac{1}{2 H^2} \int d\rho \, \frac{e^{-3\rho}}{\bar{\phi_{,\rho}^2}} \int d\rho \, e^{3\rho} \overline{V}^{(3)} \bar \phi_{,\rho}^3 \psi^2 \left(2\psi - \psi_{,\rho}\right)  = I^{(3,{\rm 4pt})} + J^{(3,{\rm 4pt})} + K^{(3,{\rm 4pt})} \,,
\end{align}
where
\begin{align}
    I^{(3,{\rm 4pt})} \equiv & - \frac{1}{2 H^2} \int d\rho \, \overline{V}^{(2)} \psi^2 \left(2 \psi - \psi_{,\rho} \right) \,, \label{eq:ZetaTilde4ptNotSimplified2} \\
    J^{(3,{\rm 4pt})} \equiv & - \frac{2}{H^4} \int d\rho \, \frac{e^{-3\rho}}{\bar \phi_{,\rho}^2 } \int d\rho \, e^{3\rho} \overline{V}^{(1)} \overline{V}^{(2)} \bar \phi_{,\rho} \psi^3   \,, \label{eq:ZetaTilde4ptNotSimplified3} \\
    K^{(3,{\rm 4pt})} \equiv & - \frac{1}{H^2} \int d\rho \, \frac{e^{-3\rho}}{\bar \phi_{,\rho}^2} \int d\rho \, e^{3\rho} \overline{V}^{(2)} \bar \phi_{,\rho}^2 \psi \left(3\psi^2 - 3 \psi \psi_{,\rho} + \psi_{,\rho}^2 \right) \,. \label{eq:ZetaTilde4ptNotSimplified4}
\end{align}
After an integration by parts in each of these terms, we find
\begin{align}
    I^{(3,{\rm 4pt})} = & \frac{\alpha}{2} \psi^2 \left( \psi_{,\rho} - 2 \psi \right) + \frac{1}{2} \int d\rho \, \left[ \alpha^2 \psi^2 \left(2\psi - 3 \psi_{,\rho} \right) + \alpha \psi \left(6 \psi^2 - 2 \psi_{,\rho}^2 \right) \right]  \,, \\
    J^{(3,{\rm 4pt})} = & - \int d\rho \, \alpha^2 \psi^3 + 3 \int d\rho \, \frac{e^{-3\rho}}{\bar \phi_{,\rho}^2 } \int d\rho \, e^{3\rho} \alpha^2 \bar \phi_{,\rho}^2 \psi^2 \left(\psi + \psi_{,\rho}\right)  \,, \\
    K^{(3,{\rm 4pt})} = & - \int d\rho \, \alpha \psi \left(3\psi^2 - 3 \psi \psi_{,\rho} + \psi_{,\rho}^2 \right) \nonumber \\
    &+ \int d\rho \, \frac{e^{-3\rho}}{\bar \phi_{,\rho}^2} \int d\rho \, e^{3\rho} \alpha \bar \phi_{,\rho} \left[ \bar \phi_{,\rho} \psi_{,\rho}^3 -3 \alpha \bar \phi_{,\rho} \psi \left(\psi^2 + \psi \psi_{,\rho} - \psi_{,\rho}^2 \right) \right]  \,.
\end{align}
All these contributions are clearly regular. This finally proves that $\tilde \zeta^{(3,{\rm 4pt})}_n$ is also regular.\\
\\
{\bf{Total contribution:}}\\
For completeness, we here present the final expression for the four-point-vertex interaction,
\begin{align}
    \tilde \zeta^{(3,{\rm 4pt})}_n 
    = & - \frac{1}{6 H^2} \overline{V}^{(2)} \psi^3 - \alpha \left(\psi- \psi_{,\rho}\right)\psi^2  +  \int d\rho \,  \left[ \alpha^2 \psi^3 - 3 \alpha^2 \psi^2 \psi_{,\rho} + \alpha \psi \psi_{,\rho}^2 \right] \nonumber \\
    & + \int d\rho \, \frac{e^{-3\rho}}{\bar \phi_{,\rho}^2} \int d\rho \, e^{3\rho} \alpha \bar \phi_{,\rho} \left[ \bar \phi_{,\rho} \psi_{,\rho}^3 - 3 \alpha \bar \phi_{,\rho} \psi\left(\psi^2 + \psi \psi_{,\rho} - \psi_{,\rho}^2 \right) \right] \,. \label{eq:ZetaTilde34pt}
\end{align}
The only contribution involving $\overline{V}^{(2)}$ is not integrated out and, as a consequence, the result remains regular when evaluated at late times.

\bibliographystyle{unsrturl}
\bibliography{deltaN}

@article{Tanaka:2026zew,
    author = "Tanaka, Takahiro and Urakawa, Yuko",
    title = "{Locality in effective field theory for inflationary soft modes}",
    eprint = "2605.19331",
    archivePrefix = "arXiv",
    primaryClass = "gr-qc",
    reportNumber = "KEK-TH-2839, KEK-Cosmo-0422",
    month = "5",
    year = "2026"
}

@article{Cruces:2025typ,
    author = "Cruces, Diego and Pi, Shi and Sasaki, Misao",
    title = "{{\ensuremath{\delta}}n formalism: A new formulation for the probability density of the curvature perturbation}",
    eprint = "2505.24590",
    archivePrefix = "arXiv",
    primaryClass = "astro-ph.CO",
    doi = "10.1103/85xd-bbgj",
    journal = "Phys. Rev. D",
    volume = "114",
    number = "2",
    pages = "023549",
    year = "2026"
}

@article{Kristiano:2025ajj,
    author = "Kristiano, Jason and Yokoyama, Jun'ichi",
    title = "{Inflationary background renormalization}",
    eprint = "2504.18514",
    archivePrefix = "arXiv",
    primaryClass = "hep-th",
    reportNumber = "YITP-25-62, RESCEU-9/25, IPMU25-0018",
    month = "4",
    year = "2025"
}

@article{Senatore:2009cf,
    author = "Senatore, Leonardo and Zaldarriaga, Matias",
    title = "{On Loops in Inflation}",
    eprint = "0912.2734",
    archivePrefix = "arXiv",
    primaryClass = "hep-th",
    doi = "10.1007/JHEP12(2010)008",
    journal = "JHEP",
    volume = "12",
    pages = "008",
    year = "2010"
}

@article{Inomata:2026csq,
    author = "Inomata, Keisuke",
    title = "{Cancellation of one-loop time dependence in superhorizon curvature perturbations from all scales}",
    eprint = "2606.28247",
    archivePrefix = "arXiv",
    primaryClass = "astro-ph.CO",
    month = "6",
    year = "2026"
}

@article{Weinberg:2003sw,
      author         = "Weinberg, Steven",
      title          = "{Adiabatic modes in cosmology}",
      journal        = "Phys. Rev.",
      volume         = "D67",
      year           = "2003",
      pages          = "123504",
      doi            = "10.1103/PhysRevD.67.123504",
      eprint         = "astro-ph/0302326",
      archivePrefix  = "arXiv",
      primaryClass   = "astro-ph",
      reportNumber   = "UTTG-12-02",
      SLACcitation   = "%%CITATION = ASTRO-PH/0302326;%%"
}

@article{Tanaka:2017nff,
      author         = "Tanaka, Takahiro and Urakawa, Yuko",
      title          = "{Large gauge transformation, Soft theorem, and Infrared
                        divergence in inflationary spacetime}",
      journal        = "JHEP",
      volume         = "10",
      year           = "2017",
      pages          = "127",
      doi            = "10.1007/JHEP10(2017)127",
      eprint         = "1707.05485",
      archivePrefix  = "arXiv",
      primaryClass   = "hep-th",
      reportNumber   = "YITP-17-76, KUNS-2694",
      SLACcitation   = "%%CITATION = ARXIV:1707.05485;%%"
}

@article{Urakawa:2010it,
    author = "Urakawa, Yuko and Tanaka, Takahiro",
    title = "{IR divergence does not affect the gauge-invariant curvature perturbation}",
    eprint = "1007.0468",
    archivePrefix = "arXiv",
    primaryClass = "hep-th",
    doi = "10.1103/PhysRevD.82.121301",
    journal = "Phys. Rev. D",
    volume = "82",
    pages = "121301",
    year = "2010"
}

@article{Urakawa:2010kr,
    author = "Urakawa, Yuko and Tanaka, Takahiro",
    title = "{Natural selection of inflationary vacuum required by infra-red regularity and gauge-invariance}",
    eprint = "1009.2947",
    archivePrefix = "arXiv",
    primaryClass = "hep-th",
    doi = "10.1143/PTP.125.1067",
    journal = "Prog. Theor. Phys.",
    volume = "125",
    pages = "1067--1089",
    year = "2011"
}

@article{Tanaka:2013xe,
    author = "Tanaka, Takahiro and Urakawa, Yuko",
    title = "{Strong restriction on inflationary vacua from the local gauge invariance II: Infrared regularity and absence of secular growth in the Euclidean vacuum}",
    eprint = "1301.3088",
    archivePrefix = "arXiv",
    primaryClass = "hep-th",
    doi = "10.1093/ptep/ptt037",
    journal = "PTEP",
    volume = "2013",
    number = "6",
    pages = "063E02",
    year = "2013"
}

@article{Tanaka:2014ina,
    author = "Tanaka, Takahiro and Urakawa, Yuko",
    title = "{Strong restriction on inflationary vacua from the local $gauge$ invariance III: Infrared regularity of graviton loops}",
    eprint = "1402.2076",
    archivePrefix = "arXiv",
    primaryClass = "hep-th",
    doi = "10.1093/ptep/ptu071",
    journal = "PTEP",
    volume = "2014",
    number = "7",
    pages = "073E01",
    year = "2014"
}

@article{Tanaka:2013caa,
    author = "Tanaka, Takahiro and Urakawa, Yuko",
    title = "{Loops in inflationary correlation functions}",
    eprint = "1306.4461",
    archivePrefix = "arXiv",
    primaryClass = "hep-th",
    doi = "10.1088/0264-9381/30/23/233001",
    journal = "Class. Quant. Grav.",
    volume = "30",
    pages = "233001",
    year = "2013"
}

@article{Tanaka:2011aj,
    author = "Tanaka, Takahiro and Urakawa, Yuko",
    title = "{Dominance of gauge artifact in the consistency relation for the primordial bispectrum}",
    eprint = "1103.1251",
    archivePrefix = "arXiv",
    primaryClass = "astro-ph.CO",
    reportNumber = "YITP-11-31",
    doi = "10.1088/1475-7516/2011/05/014",
    journal = "JCAP",
    volume = "05",
    pages = "014",
    year = "2011"
}

@article{Pajer:2013ana,
    author = "Pajer, Enrico and Schmidt, Fabian and Zaldarriaga, Matias",
    title = "{The Observed Squeezed Limit of Cosmological Three-Point Functions}",
    eprint = "1305.0824",
    archivePrefix = "arXiv",
    primaryClass = "astro-ph.CO",
    doi = "10.1103/PhysRevD.88.083502",
    journal = "Phys. Rev. D",
    volume = "88",
    number = "8",
    pages = "083502",
    year = "2013"
}

@book{Byrnes:2025tji,
    editor = "Byrnes, Christian and Franciolini, Gabriele and Harada, Tomohiro and Pani, Paolo and Sasaki, Misao",
    title = "{Primordial Black Holes}",
    doi = "10.1007/978-981-97-8887-3",
    isbn = "978-981--978886-6, 978-981--978889-7, 978-981--978887-3",
    publisher = "Springer",
    series = "Springer Series in Astrophysics and Cosmology",
    year = "2025"
}

@article{Escriva:2022duf,
    author = "Escriv{\`a}, Albert and Kuhnel, Florian and Tada, Yuichiro",
    editor = "Sedda, Manuel Arca and Bortolas, Elisa and Spera, Mario",
    title = "{Primordial Black Holes}",
    eprint = "2211.05767",
    archivePrefix = "arXiv",
    primaryClass = "astro-ph.CO",
    doi = "10.1016/B978-0-32-395636-9.00012-8",
    month = "11",
    year = "2022"
}

@article{Escriva:2025ftp,
    author = "Escriv{\`a}, Albert and Garriga, Jaume and Pi, Shi",
    title = "{Inflationary relics from an ultra-slow-roll plateau}",
    eprint = "2512.04986",
    archivePrefix = "arXiv",
    primaryClass = "astro-ph.CO",
    doi = "10.1088/1475-7516/2026/03/018",
    journal = "JCAP",
    volume = "03",
    pages = "018",
    year = "2026"
}

@article{Ahmadi:2026yrv,
    author = "Ahmadi, S. Mohammad",
    title = "{Continuous Sensitivity Analysis for $\delta N$ Formalism}",
    eprint = "2603.27366",
    archivePrefix = "arXiv",
    primaryClass = "gr-qc",
    month = "3",
    year = "2026"
}

@article{Ahmadi:2026rzf,
    author = "Ahmadi, S. Mohammad and Ahmadi, Nahid",
    title = "{$\delta N$ formalism with gradient interactions}",
    eprint = "2602.00902",
    archivePrefix = "arXiv",
    primaryClass = "gr-qc",
    month = "1",
    year = "2026"
}

@article{Braglia:2026fle,
    author = "Braglia, Matteo and C{\'e}spedes, Sebasti{\'a}n and Pinol, Lucas",
    title = "{Scale-Dependent Loop Corrections to the Inflationary Power Spectrum}",
    eprint = "2603.12216",
    archivePrefix = "arXiv",
    primaryClass = "astro-ph.CO",
    reportNumber = "CERN-TH-2026-038",
    month = "3",
    year = "2026"
}

@article{Braglia:2025qrb,
    author = "Braglia, Matteo and Pinol, Lucas",
    title = "{Freezing of the renormalized one-loop primordial scalar power spectrum}",
    eprint = "2504.13136",
    archivePrefix = "arXiv",
    primaryClass = "astro-ph.CO",
    doi = "10.1103/5jkv-mj3w",
    journal = "Phys. Rev. D",
    volume = "113",
    number = "6",
    pages = "L061302",
    year = "2026"
}

@article{Braglia:2025cee,
    author = "Braglia, Matteo and Pinol, Lucas",
    title = "{One-loop renormalization of the effective field theory of inflationary fluctuations from gravitational interactions}",
    eprint = "2504.07926",
    archivePrefix = "arXiv",
    primaryClass = "astro-ph.CO",
    doi = "10.1103/f6q6-5jxb",
    journal = "Phys. Rev. D",
    volume = "113",
    number = "6",
    pages = "063513",
    year = "2026"
}

@article{Yang:1950vi,
    author = "Yang, Chen-Ning and Feldman, D.",
    title = "{The S Matrix in the Heisenberg Representation}",
    doi = "10.1103/PhysRev.79.972",
    journal = "Phys. Rev.",
    volume = "79",
    pages = "972--978",
    year = "1950"
}

@article{Arroja:2011yj,
    author = "Arroja, Frederico and Tanaka, Takahiro",
    title = "{A note on the role of the boundary terms for the non-Gaussianity in general k-inflation}",
    eprint = "1103.1102",
    archivePrefix = "arXiv",
    primaryClass = "astro-ph.CO",
    doi = "10.1088/1475-7516/2011/05/005",
    journal = "JCAP",
    volume = "05",
    pages = "005",
    year = "2011"
}

@article{Kawaguchi:2024lsw,
    author = "Kawaguchi, Ryodai and Tsujikawa, Shinji and Yamada, Yusuke",
    title = "{Roles of boundary and equation-of-motion terms in cosmological correlation functions}",
    eprint = "2403.16022",
    archivePrefix = "arXiv",
    primaryClass = "hep-th",
    reportNumber = "WUCG-24-03",
    doi = "10.1016/j.physletb.2024.138962",
    journal = "Phys. Lett. B",
    volume = "856",
    pages = "138962",
    year = "2024"
}

@article{Dimopoulos:2017ged,
    author = "Dimopoulos, Konstantinos",
    title = "{Ultra slow-roll inflation demystified}",
    eprint = "1707.05644",
    archivePrefix = "arXiv",
    primaryClass = "hep-ph",
    doi = "10.1016/j.physletb.2017.10.066",
    journal = "Phys. Lett. B",
    volume = "775",
    pages = "262--265",
    year = "2017"
}

@article{Cruces:2026qvl,
    author = "Cruces, Diego and He, Minxi and Pi, Shi and Wang, Jianing and Yamaguchi, Masahide and Zhu, Yuhang",
    title = "{Natura Non Facit Saltum: An Analytical Model of Smooth Slow-Roll to Ultra-Slow-Roll Transition}",
    eprint = "2603.17465",
    archivePrefix = "arXiv",
    primaryClass = "astro-ph.CO",
    month = "3",
    year = "2026"
}

@article{Inoue:2001zt,
    author = "Inoue, Shogo and Yokoyama, Jun'ichi",
    title = "{Curvature perturbation at the local extremum of the inflaton's potential}",
    eprint = "hep-ph/0104083",
    archivePrefix = "arXiv",
    reportNumber = "OU-TAP-160",
    doi = "10.1016/S0370-2693(01)01369-7",
    journal = "Phys. Lett. B",
    volume = "524",
    pages = "15--20",
    year = "2002"
}

@inproceedings{Iacconi:2026vyk,
    author = "Iacconi, Laura",
    title = "{Primordial black holes from inflation: on the decoupling between large and small scales}",
    booktitle = "{60th Rencontres de Moriond on Cosmology}: {Moriond Cosmology 2026}",
    eprint = "2605.04815",
    archivePrefix = "arXiv",
    primaryClass = "astro-ph.CO",
    month = "5",
    year = "2026"
}

@article{Ema:2026dop,
    author = "Ema, Yohei and Hong, Muzi and Jinno, Ryusuke and Mukaida, Kyohei",
    title = "{Cancellation of loop corrections to soft scalar power spectrum}",
    eprint = "2603.01961",
    archivePrefix = "arXiv",
    primaryClass = "astro-ph.CO",
    month = "3",
    year = "2026"
}

@article{Kawaguchi:2024rsv,
    author = "Kawaguchi, Ryodai and Tsujikawa, Shinji and Yamada, Yusuke",
    title = "{Proving the absence of large one-loop corrections to the power spectrum of curvature perturbations in transient ultra-slow-roll inflation within the path-integral approach}",
    eprint = "2407.19742",
    archivePrefix = "arXiv",
    primaryClass = "hep-th",
    reportNumber = "WUCG-24-07",
    doi = "10.1007/JHEP12(2024)095",
    journal = "JHEP",
    volume = "12",
    pages = "095",
    year = "2024"
}

@article{Iacconi:2026uzo,
    author = "Iacconi, Laura and Mulryne, David and Seery, David",
    title = "{Decoupling of large-scale, adiabatic inflationary perturbations from enhanced small-scale modes at one-loop}",
    eprint = "2601.14229",
    archivePrefix = "arXiv",
    primaryClass = "astro-ph.CO",
    month = "1",
    year = "2026"
}

@article{Fumagalli:2024jzz,
    author = "Fumagalli, Jacopo",
    title = "{Absence of one-loop effects on large scales from small scales in non-slow-roll dynamics. Part 2. Quartic interactions and consistency relations}",
    eprint = "2408.08296",
    archivePrefix = "arXiv",
    primaryClass = "astro-ph.CO",
    doi = "10.1007/JHEP01(2025)108",
    journal = "JHEP",
    volume = "01",
    pages = "108",
    year = "2025"
}

@article{Choudhury:2023hvf,
    author = "Choudhury, Sayantan and Panda, Sudhakar and Sami, M.",
    title = "{Galileon inflation evades the no-go for PBH formation in the single-field framework}",
    eprint = "2304.04065",
    archivePrefix = "arXiv",
    primaryClass = "astro-ph.CO",
    doi = "10.1088/1475-7516/2023/08/078",
    journal = "JCAP",
    volume = "08",
    pages = "078",
    year = "2023"
}

@article{Choudhury:2023jlt,
    author = "Choudhury, Sayantan and Panda, Sudhakar and Sami, M.",
    title = "{PBH formation in EFT of single field inflation with sharp transition}",
    eprint = "2302.05655",
    archivePrefix = "arXiv",
    primaryClass = "astro-ph.CO",
    doi = "10.1016/j.physletb.2023.138123",
    journal = "Phys. Lett. B",
    volume = "845",
    pages = "138123",
    year = "2023"
}

@article{Riotto:2023hoz,
    author = "Riotto, Antonio",
    title = "{The Primordial Black Hole Formation from Single-Field Inflation is Not Ruled Out}",
    eprint = "2301.00599",
    archivePrefix = "arXiv",
    primaryClass = "astro-ph.CO",
    month = "1",
    year = "2023"
}

@article{Choudhury:2023vuj,
    author = "Choudhury, Sayantan and Gangopadhyay, Mayukh R. and Sami, M.",
    title = "{No-go for the formation of heavy mass Primordial Black Holes in Single Field Inflation}",
    eprint = "2301.10000",
    archivePrefix = "arXiv",
    primaryClass = "astro-ph.CO",
    doi = "10.1140/epjc/s10052-024-13218-2",
    journal = "Eur. Phys. J. C",
    volume = "84",
    number = "9",
    pages = "884",
    year = "2024"
}

@article{Malik:2008im,
    author = "Malik, Karim A. and Wands, David",
    title = "{Cosmological perturbations}",
    eprint = "0809.4944",
    archivePrefix = "arXiv",
    primaryClass = "astro-ph",
    doi = "10.1016/j.physrep.2009.03.001",
    journal = "Phys. Rept.",
    volume = "475",
    pages = "1--51",
    year = "2009"
}

@article{Pattison:2019hef,
    author = "Pattison, Chris and Vennin, Vincent and Assadullahi, Hooshyar and Wands, David",
    title = "{Stochastic inflation beyond slow roll}",
    eprint = "1905.06300",
    archivePrefix = "arXiv",
    primaryClass = "astro-ph.CO",
    doi = "10.1088/1475-7516/2019/07/031",
    journal = "JCAP",
    volume = "07",
    pages = "031",
    year = "2019"
}

@article{Briaud:2025ayt,
    author = "Briaud, Vadim and Kawaguchi, Ryodai and Vennin, Vincent",
    title = "{Stochastic inflation with gradient interactions}",
    eprint = "2509.05124",
    archivePrefix = "arXiv",
    primaryClass = "astro-ph.CO",
    doi = "10.1088/1475-7516/2025/12/024",
    journal = "JCAP",
    volume = "12",
    pages = "024",
    year = "2025"
}

@article{Pattison:2018bct,
    author = "Pattison, Chris and Vennin, Vincent and Assadullahi, Hooshyar and Wands, David",
    title = "{The attractive behaviour of ultra-slow-roll inflation}",
    eprint = "1806.09553",
    archivePrefix = "arXiv",
    primaryClass = "astro-ph.CO",
    doi = "10.1088/1475-7516/2018/08/048",
    journal = "JCAP",
    volume = "08",
    pages = "048",
    year = "2018"
}

@article{Starobinsky:1982ee,
      author         = "Starobinsky, Alexei A.",
      title          = "{Dynamics of Phase Transition in the New Inflationary
                        Universe Scenario and Generation of Perturbations}",
      journal        = "Phys. Lett.",
      volume         = "117B",
      year           = "1982",
      pages          = "175-178",
      doi            = "10.1016/0370-2693(82)90541-X",
      SLACcitation   = "%%CITATION = PHLTA,117B,175;%%"
}

@article{Starobinsky:1992ts,
    author = "Starobinsky, Alexei A.",
    title = "{Spectrum of adiabatic perturbations in the universe when there are singularities in the inflation potential}",
    journal = "JETP Lett.",
    volume = "55",
    pages = "489--494",
    year = "1992"
}

@article{Salopek:1990jq,
    author = "Salopek, D. S. and Bond, J. R.",
    title = "{Nonlinear evolution of long wavelength metric fluctuations in inflationary models}",
    reportNumber = "FERMILAB-PUB-90-131-A",
    doi = "10.1103/PhysRevD.42.3936",
    journal = "Phys. Rev. D",
    volume = "42",
    pages = "3936--3962",
    year = "1990"
}

@article{Sasaki:1995aw,
    author = "Sasaki, Misao and Stewart, Ewan D.",
    title = "{A General analytic formula for the spectral index of the density perturbations produced during inflation}",
    eprint = "astro-ph/9507001",
    archivePrefix = "arXiv",
    reportNumber = "LANCS-TH-9504, OU-TAP-22",
    doi = "10.1143/PTP.95.71",
    journal = "Prog. Theor. Phys.",
    volume = "95",
    pages = "71--78",
    year = "1996"
}

@article{Wands:2000dp,
    author = "Wands, David and Malik, Karim A. and Lyth, David H. and Liddle, Andrew R.",
    title = "{A New approach to the evolution of cosmological perturbations on large scales}",
    eprint = "astro-ph/0003278",
    archivePrefix = "arXiv",
    doi = "10.1103/PhysRevD.62.043527",
    journal = "Phys. Rev. D",
    volume = "62",
    pages = "043527",
    year = "2000"
}

@article{Lyth:2003im,
    author = "Lyth, David H. and Wands, David",
    title = "{Conserved cosmological perturbations}",
    eprint = "astro-ph/0306498",
    archivePrefix = "arXiv",
    doi = "10.1103/PhysRevD.68.103515",
    journal = "Phys. Rev. D",
    volume = "68",
    pages = "103515",
    year = "2003"
}

@article{Rigopoulos:2003ak,
    author = "Rigopoulos, G. I. and Shellard, E. P. S.",
    title = "{The separate universe approach and the evolution of nonlinear superhorizon cosmological perturbations}",
    eprint = "astro-ph/0306620",
    archivePrefix = "arXiv",
    doi = "10.1103/PhysRevD.68.123518",
    journal = "Phys. Rev. D",
    volume = "68",
    pages = "123518",
    year = "2003"
}

@article{Lyth:2004gb,
    author = "Lyth, David H. and Malik, Karim A. and Sasaki, Misao",
    title = "{A General proof of the conservation of the curvature perturbation}",
    eprint = "astro-ph/0411220",
    archivePrefix = "arXiv",
    reportNumber = "YITP-04-67",
    doi = "10.1088/1475-7516/2005/05/004",
    journal = "JCAP",
    volume = "05",
    pages = "004",
    year = "2005"
}

@article{Lyth:2005fi,
    author = "Lyth, David H. and Rodriguez, Yeinzon",
    title = "{The Inflationary prediction for primordial non-Gaussianity}",
    eprint = "astro-ph/0504045",
    archivePrefix = "arXiv",
    doi = "10.1103/PhysRevLett.95.121302",
    journal = "Phys. Rev. Lett.",
    volume = "95",
    pages = "121302",
    year = "2005"
}

@article{Abolhasani:2013zya,
    author = "Abolhasani, Ali Akbar and Emami, Razieh and Firouzjaee, Javad T. and Firouzjahi, Hassan",
    title = "{$\delta N$ formalism in anisotropic inflation and large anisotropic bispectrum and trispectrum}",
    eprint = "1302.6986",
    archivePrefix = "arXiv",
    primaryClass = "astro-ph.CO",
    doi = "10.1088/1475-7516/2013/08/016",
    journal = "JCAP",
    volume = "08",
    pages = "016",
    year = "2013"
}

@article{Talebian-Ashkezari:2016llx,
    author = "Talebian-Ashkezari, Alireza and Ahmadi, Nahid and Abolhasani, Ali Akbar",
    title = "{$\delta M$ formalism: a new approach to cosmological perturbation theory in anisotropic inflation}",
    eprint = "1609.05893",
    archivePrefix = "arXiv",
    primaryClass = "gr-qc",
    doi = "10.1088/1475-7516/2018/03/001",
    journal = "JCAP",
    volume = "03",
    pages = "001",
    year = "2018"
}

@article{Talebian-Ashkezari:2018cax,
    author = "Talebian-Ashkezari, Alireza and Ahmadi, Nahid",
    title = "{$\delta M$formalism and anisotropic chaotic inflation power spectrum}",
    eprint = "1803.03763",
    archivePrefix = "arXiv",
    primaryClass = "gr-qc",
    doi = "10.1088/1475-7516/2018/05/047",
    journal = "JCAP",
    volume = "05",
    pages = "047",
    year = "2018"
}

@article{Tanaka:2021dww,
    author = "Tanaka, Takahiro and Urakawa, Yuko",
    title = "{Anisotropic separate universe and Weinberg's adiabatic mode}",
    eprint = "2101.05707",
    archivePrefix = "arXiv",
    primaryClass = "astro-ph.CO",
    doi = "10.1088/1475-7516/2021/07/051",
    journal = "JCAP",
    volume = "07",
    pages = "051",
    year = "2021"
}

@article{Tanaka:2023gul,
    author = "Tanaka, Takahiro and Urakawa, Yuko",
    title = "{Statistical Anisotropy of Primordial Gravitational Waves from Generalized {\ensuremath{\delta}}N Formalism}",
    eprint = "2309.08497",
    archivePrefix = "arXiv",
    primaryClass = "gr-qc",
    reportNumber = "KUNS-2978",
    doi = "10.1103/PhysRevLett.132.231003",
    journal = "Phys. Rev. Lett.",
    volume = "132",
    number = "23",
    pages = "231003",
    year = "2024"
}

@article{Sasaki:1998ug,
    author = "Sasaki, Misao and Tanaka, Takahiro",
    title = "{Superhorizon scale dynamics of multiscalar inflation}",
    eprint = "gr-qc/9801017",
    archivePrefix = "arXiv",
    reportNumber = "OU-TAP-72",
    doi = "10.1143/PTP.99.763",
    journal = "Prog. Theor. Phys.",
    volume = "99",
    pages = "763--782",
    year = "1998"
}

@article{Maldacena:2002vr,
    author = "Maldacena, Juan Martin",
    title = "{Non-Gaussian features of primordial fluctuations in single field inflationary models}",
    eprint = "astro-ph/0210603",
    archivePrefix = "arXiv",
    doi = "10.1088/1126-6708/2003/05/013",
    journal = "JHEP",
    volume = "05",
    pages = "013",
    year = "2003"
}

@article{Artigas:2021zdk,
    author = "Artigas, Danilo and Grain, Julien and Vennin, Vincent",
    title = "{Hamiltonian formalism for cosmological perturbations: the~separate-universe approach}",
    eprint = "2110.11720",
    archivePrefix = "arXiv",
    primaryClass = "astro-ph.CO",
    doi = "10.1088/1475-7516/2022/02/001",
    journal = "JCAP",
    volume = "02",
    number = "02",
    pages = "001",
    year = "2022"
}

@article{Naruko:2012fe,
    author = "Naruko, Atsushi and Takamizu, Yu-ichi and Sasaki, Misao",
    title = "{Beyond $\delta N$ formalism}",
    eprint = "1210.6525",
    archivePrefix = "arXiv",
    primaryClass = "astro-ph.CO",
    reportNumber = "YITP-12-75",
    doi = "10.1093/ptep/ptt008",
    journal = "PTEP",
    volume = "2013",
    pages = "043E01",
    year = "2013"
}

@article{Leach:2001zf,
    author = "Leach, Samuel M and Sasaki, Misao and Wands, David and Liddle, Andrew R",
    title = "{Enhancement of superhorizon scale inflationary curvature perturbations}",
    eprint = "astro-ph/0101406",
    archivePrefix = "arXiv",
    doi = "10.1103/PhysRevD.64.023512",
    journal = "Phys. Rev. D",
    volume = "64",
    pages = "023512",
    year = "2001"
}

@article{Jackson:2023obv,
    author = "Jackson, Joseph H. P. and Assadullahi, Hooshyar and Gow, Andrew D. and Koyama, Kazuya and Vennin, Vincent and Wands, David",
    title = "{The separate-universe approach and sudden transitions during inflation}",
    eprint = "2311.03281",
    archivePrefix = "arXiv",
    primaryClass = "astro-ph.CO",
    doi = "10.1088/1475-7516/2024/05/053",
    journal = "JCAP",
    volume = "05",
    pages = "053",
    year = "2024"
}

@inbook{Kristiano:2024ngc,
    author = "Kristiano, Jason and Yokoyama, Jun'ichi",
    title = "{Generating Large Primordial Fluctuations in Single-Field Inflation for~Primordial Black Hole Formation}",
    eprint = "2405.12149",
    archivePrefix = "arXiv",
    primaryClass = "astro-ph.CO",
    reportNumber = "RESCEU-7/24",
    doi = "10.1007/978-981-97-8887-3_3",
    year = "2025"
}

@article{Kristiano:2024vst,
    author = "Kristiano, Jason and Yokoyama, Jun'ichi",
    title = "{Comparing sharp and smooth transitions of the second slow-roll parameter in single-field inflation}",
    eprint = "2405.12145",
    archivePrefix = "arXiv",
    primaryClass = "astro-ph.CO",
    reportNumber = "RESCEU-8/24",
    doi = "10.1088/1475-7516/2024/10/036",
    journal = "JCAP",
    volume = "10",
    pages = "036",
    year = "2024"
}

@article{Davies:2023hhn,
    author = "Davies, Matthew W. and Iacconi, Laura and Mulryne, David J.",
    title = "{Numerical 1-loop correction from a potential yielding ultra-slow-roll dynamics}",
    eprint = "2312.05694",
    archivePrefix = "arXiv",
    primaryClass = "astro-ph.CO",
    doi = "10.1088/1475-7516/2024/04/050",
    journal = "JCAP",
    volume = "04",
    pages = "050",
    year = "2024"
}

@article{Tasinato:2023ukp,
    author = "Tasinato, Gianmassimo",
    title = "{Large $\eta$ approach to single field inflation}",
    eprint = "2305.11568",
    archivePrefix = "arXiv",
    primaryClass = "hep-th",
    doi = "10.1103/PhysRevD.108.043526",
    journal = "Phys. Rev. D",
    volume = "108",
    number = "4",
    pages = "043526",
    year = "2023"
}

@article{Tada:2023rgp,
    author = "Tada, Yuichiro and Terada, Takahiro and Tokuda, Junsei",
    title = "{Cancellation of quantum corrections on the soft curvature perturbations}",
    eprint = "2308.04732",
    archivePrefix = "arXiv",
    primaryClass = "hep-th",
    reportNumber = "CTPU-PTC-23-31",
    doi = "10.1007/JHEP01(2024)105",
    journal = "JHEP",
    volume = "01",
    pages = "105",
    year = "2024"
}

@article{Ballesteros:2024zdp,
    author = "Ballesteros, Guillermo and Gamb{\'\i}n Egea, Jes{\'u}s",
    title = "{One-loop power spectrum in ultra slow-roll inflation and implications for primordial black hole dark matter}",
    eprint = "2404.07196",
    archivePrefix = "arXiv",
    primaryClass = "astro-ph.CO",
    doi = "10.1088/1475-7516/2024/07/052",
    journal = "JCAP",
    volume = "07",
    pages = "052",
    year = "2024"
}

@article{Inomata:2025pqa,
    author = "Inomata, Keisuke",
    title = "{Role of the counterterms in the conservation of superhorizon curvature perturbations at one loop}",
    eprint = "2502.12112",
    archivePrefix = "arXiv",
    primaryClass = "astro-ph.CO",
    doi = "10.1103/r8bh-s48f",
    journal = "Phys. Rev. D",
    volume = "111",
    number = "12",
    pages = "123517",
    year = "2025"
}

@article{Inomata:2025bqw,
    author = "Inomata, Keisuke",
    title = "{Conservation of superhorizon curvature perturbations at one loop: Backreaction in the in-in formalism and renormalization}",
    eprint = "2502.08707",
    archivePrefix = "arXiv",
    primaryClass = "astro-ph.CO",
    doi = "10.1103/PhysRevD.111.103504",
    journal = "Phys. Rev. D",
    volume = "111",
    number = "10",
    pages = "103504",
    year = "2025"
}

@article{Inomata:2024lud,
    author = "Inomata, Keisuke",
    title = "{Superhorizon Curvature Perturbations Are Protected against One-Loop Corrections}",
    eprint = "2403.04682",
    archivePrefix = "arXiv",
    primaryClass = "astro-ph.CO",
    doi = "10.1103/PhysRevLett.133.141001",
    journal = "Phys. Rev. Lett.",
    volume = "133",
    number = "14",
    pages = "141001",
    year = "2024"
}

@article{Choudhury:2023rks,
    author = "Choudhury, Sayantan and Panda, Sudhakar and Sami, M.",
    title = "{Quantum loop effects on the power spectrum and constraints on primordial black holes}",
    eprint = "2303.06066",
    archivePrefix = "arXiv",
    primaryClass = "astro-ph.CO",
    doi = "10.1088/1475-7516/2023/11/066",
    journal = "JCAP",
    volume = "11",
    pages = "066",
    year = "2023"
}

@article{Caravano:2025diq,
    author = "Caravano, Angelo and Franciolini, Gabriele and Renaux-Petel, S{\'e}bastien",
    title = "{Ultraslow-roll inflation on the lattice. II. Nonperturbative curvature perturbation}",
    eprint = "2506.11795",
    archivePrefix = "arXiv",
    primaryClass = "astro-ph.CO",
    reportNumber = "CERN-TH-2025-116",
    doi = "10.1103/39qd-gdfm",
    journal = "Phys. Rev. D",
    volume = "112",
    number = "8",
    pages = "083508",
    year = "2025"
}

@article{Caravano:2024moy,
    author = "Caravano, Angelo and Franciolini, Gabriele and Renaux-Petel, S{\'e}bastien",
    title = "{Ultraslow-roll inflation on the lattice: Backreaction and nonlinear effects}",
    eprint = "2410.23942",
    archivePrefix = "arXiv",
    primaryClass = "astro-ph.CO",
    reportNumber = "CERN-TH-2024-181",
    doi = "10.1103/PhysRevD.111.063518",
    journal = "Phys. Rev. D",
    volume = "111",
    number = "6",
    pages = "063518",
    year = "2025"
}

@article{Pi:2022zxs,
    author = "Pi, Shi and Wang, Jianing",
    title = "{Primordial black hole formation in Starobinsky's linear potential model}",
    eprint = "2209.14183",
    archivePrefix = "arXiv",
    primaryClass = "astro-ph.CO",
    reportNumber = "IPMU22-0047",
    doi = "10.1088/1475-7516/2023/06/018",
    journal = "JCAP",
    volume = "06",
    pages = "018",
    year = "2023"
}

@article{Franciolini:2023agm,
    author = "Franciolini, Gabriele and Iovino, Junior., Antonio and Taoso, Marco and Urbano, Alfredo",
    title = "{Perturbativity in the presence of ultraslow-roll dynamics}",
    eprint = "2305.03491",
    archivePrefix = "arXiv",
    primaryClass = "astro-ph.CO",
    doi = "10.1103/PhysRevD.109.123550",
    journal = "Phys. Rev. D",
    volume = "109",
    number = "12",
    pages = "123550",
    year = "2024"
}

@article{Iacconi:2023ggt,
    author = "Iacconi, Laura and Mulryne, David and Seery, David",
    title = "{Loop corrections in the separate universe picture}",
    eprint = "2312.12424",
    archivePrefix = "arXiv",
    primaryClass = "astro-ph.CO",
    doi = "10.1088/1475-7516/2024/06/062",
    journal = "JCAP",
    volume = "06",
    pages = "062",
    year = "2024"
}

@article{Fumagalli:2023zzl,
    author = "Fumagalli, Jacopo",
    title = "{Absence of one-loop effects on large scales from small scales in non-slow-roll dynamics}",
    eprint = "2305.19263",
    archivePrefix = "arXiv",
    primaryClass = "astro-ph.CO",
    doi = "10.1007/JHEP05(2025)162",
    journal = "JHEP",
    volume = "05",
    pages = "162",
    year = "2025"
}

@article{Motohashi:2023syh,
    author = "Motohashi, Hayato and Tada, Yuichiro",
    title = "{Squeezed bispectrum and one-loop corrections in transient constant-roll inflation}",
    eprint = "2303.16035",
    archivePrefix = "arXiv",
    primaryClass = "astro-ph.CO",
    doi = "10.1088/1475-7516/2023/08/069",
    journal = "JCAP",
    volume = "08",
    pages = "069",
    year = "2023"
}

@article{Maity:2023qzw,
    author = "Maity, Suvashis and Ragavendra, H. V. and Sethi, Shiv K. and Sriramkumar, L.",
    title = "{Loop contributions to the scalar power spectrum due to quartic order action in ultra slow roll inflation}",
    eprint = "2307.13636",
    archivePrefix = "arXiv",
    primaryClass = "astro-ph.CO",
    doi = "10.1088/1475-7516/2024/05/046",
    journal = "JCAP",
    volume = "05",
    pages = "046",
    year = "2024"
}

@article{Cheng:2023ikq,
    author = "Cheng, Shu-Lin and Lee, Da-Shin and Ng, Kin-Wang",
    title = "{Primordial perturbations from ultra-slow-roll single-field inflation with quantum loop effects}",
    eprint = "2305.16810",
    archivePrefix = "arXiv",
    primaryClass = "astro-ph.CO",
    doi = "10.1088/1475-7516/2024/03/008",
    journal = "JCAP",
    volume = "03",
    pages = "008",
    year = "2024"
}

@article{Kristiano:2023scm,
    author = "Kristiano, Jason and Yokoyama, Jun'ichi",
    title = "{Note on the bispectrum and one-loop corrections in single-field inflation with primordial black hole formation}",
    eprint = "2303.00341",
    archivePrefix = "arXiv",
    primaryClass = "hep-th",
    reportNumber = "RESCEU-3/23",
    doi = "10.1103/PhysRevD.109.103541",
    journal = "Phys. Rev. D",
    volume = "109",
    number = "10",
    pages = "103541",
    year = "2024"
}

@article{Firouzjahi:2023aum,
    author = "Firouzjahi, Hassan",
    title = "{One-loop corrections in power spectrum in single field inflation}",
    eprint = "2303.12025",
    archivePrefix = "arXiv",
    primaryClass = "astro-ph.CO",
    doi = "10.1088/1475-7516/2023/10/006",
    journal = "JCAP",
    volume = "10",
    pages = "006",
    year = "2023"
}

@article{Riotto:2023gpm,
    author = "Riotto, A.",
    title = "{The Primordial Black Hole Formation from Single-Field Inflation is Still Not Ruled Out}",
    eprint = "2303.01727",
    archivePrefix = "arXiv",
    primaryClass = "astro-ph.CO",
    month = "3",
    year = "2023"
}

@article{Firouzjahi:2023ahg,
    author = "Firouzjahi, Hassan and Riotto, Antonio",
    title = "{Primordial Black Holes and loops in single-field inflation}",
    eprint = "2304.07801",
    archivePrefix = "arXiv",
    primaryClass = "astro-ph.CO",
    doi = "10.1088/1475-7516/2024/02/021",
    journal = "JCAP",
    volume = "02",
    pages = "021",
    year = "2024"
}

@article{Kristiano:2022maq,
    author = "Kristiano, Jason and Yokoyama, Jun'ichi",
    title = "{Constraining Primordial Black Hole Formation from Single-Field Inflation}",
    eprint = "2211.03395",
    archivePrefix = "arXiv",
    primaryClass = "hep-th",
    reportNumber = "RESCEU-20/22",
    doi = "10.1103/PhysRevLett.132.221003",
    journal = "Phys. Rev. Lett.",
    volume = "132",
    number = "22",
    pages = "221003",
    year = "2024"
}

@article{Martin:2012pe,
    author = "Martin, Jerome and Motohashi, Hayato and Suyama, Teruaki",
    title = "{Ultra Slow-Roll Inflation and the non-Gaussianity Consistency Relation}",
    eprint = "1211.0083",
    archivePrefix = "arXiv",
    primaryClass = "astro-ph.CO",
    reportNumber = "RESCEU-47-12",
    doi = "10.1103/PhysRevD.87.023514",
    journal = "Phys. Rev. D",
    volume = "87",
    number = "2",
    pages = "023514",
    year = "2013"
}

@article{Carr:2016drx,
    author = "Carr, Bernard and Kuhnel, Florian and Sandstad, Marit",
    title = "{Primordial Black Holes as Dark Matter}",
    eprint = "1607.06077",
    archivePrefix = "arXiv",
    primaryClass = "astro-ph.CO",
    reportNumber = "NORDITA-2016-83",
    doi = "10.1103/PhysRevD.94.083504",
    journal = "Phys. Rev. D",
    volume = "94",
    number = "8",
    pages = "083504",
    year = "2016"
}

@article{Carr:2020xqk,
	author = "Carr, Bernard and Kuhnel, Florian",
	title = "{Primordial Black Holes as Dark Matter: Recent Developments}",
	eprint = "2006.02838",
	archivePrefix = "arXiv",
	primaryClass = "astro-ph.CO",
	doi = "10.1146/annurev-nucl-050520-125911",
	journal = "Ann. Rev. Nucl. Part. Sci.",
	volume = "70",
	pages = "355--394",
	year = "2020"
}

@article{Atal:2019cdz,
	author = "Atal, Vicente and Garriga, Jaume and Marcos-Caballero, Airam",
	title = "{Primordial black hole formation with non-Gaussian curvature perturbations}",
	eprint = "1905.13202",
	archivePrefix = "arXiv",
	primaryClass = "astro-ph.CO",
	doi = "10.1088/1475-7516/2019/09/073",
	journal = "JCAP",
	volume = "09",
	pages = "073",
	year = "2019"
}

@article{Biagetti:2018pjj,
    author = "Biagetti, Matteo and Franciolini, Gabriele and Kehagias, Alex and Riotto, Antonio",
    title = "{Primordial Black Holes from Inflation and Quantum Diffusion}",
    eprint = "1804.07124",
    archivePrefix = "arXiv",
    primaryClass = "astro-ph.CO",
    doi = "10.1088/1475-7516/2018/07/032",
    journal = "JCAP",
    volume = "07",
    pages = "032",
    year = "2018"
}

@article{Kinney:2005vj,
    author = "Kinney, William H.",
    title = "{Horizon crossing and inflation with large eta}",
    eprint = "gr-qc/0503017",
    archivePrefix = "arXiv",
    doi = "10.1103/PhysRevD.72.023515",
    journal = "Phys. Rev. D",
    volume = "72",
    pages = "023515",
    year = "2005"
}

@article{Leach:2000yw,
    author = "Leach, Samuel M. and Liddle, Andrew R.",
    title = "{Inflationary perturbations near horizon crossing}",
    eprint = "astro-ph/0010082",
    archivePrefix = "arXiv",
    doi = "10.1103/PhysRevD.63.043508",
    journal = "Phys. Rev. D",
    volume = "63",
    pages = "043508",
    year = "2001"
}

@Article{Pi:2022ysn,
  author        = {Pi, Shi and Sasaki, Misao},
  journal       = {Phys. Rev. Lett.},
  title         = {{Logarithmic Duality of the Curvature Perturbation}},
  year          = {2023},
  number        = {1},
  pages         = {011002},
  volume        = {131},
  archiveprefix = {arXiv},
  doi           = {10.1103/PhysRevLett.131.011002},
  eprint        = {2211.13932},
  primaryclass  = {astro-ph.CO},
  reportnumber  = {IPMU22-0060, YITP-22-144},
}

@article{Green:2020jor,
	author = "Green, Anne M. and Kavanagh, Bradley J.",
	title = "{Primordial Black Holes as a dark matter candidate}",
	eprint = "2007.10722",
	archivePrefix = "arXiv",
	primaryClass = "astro-ph.CO",
	month = "7",
	year = "2020"
}

@article{Artigas:2024ajh,
    author = "Artigas, Danilo and Pi, Shi and Tanaka, Takahiro",
    title = "{Extended {\ensuremath{\delta}}N Formalism: Nonspatially Flat Separate-Universe Approach}",
    eprint = "2408.09964",
    archivePrefix = "arXiv",
    primaryClass = "astro-ph.CO",
    doi = "10.1103/PhysRevLett.134.221001",
    journal = "Phys. Rev. Lett.",
    volume = "134",
    number = "22",
    pages = "221001",
    year = "2025"
}

\end{document}